\documentclass{article}

\usepackage{arxiv}
\usepackage{xcolor} 
\usepackage{graphicx}
\usepackage{natbib}
\usepackage{subcaption} 
\setcitestyle{authoryear}
\usepackage{amsmath} 
\usepackage{booktabs} 

\usepackage[utf8]{inputenc} 
\usepackage[T1]{fontenc}    
\usepackage{hyperref}       
\usepackage{url}            
\usepackage{booktabs}       
\usepackage{amsfonts}       
\usepackage{nicefrac}       
\usepackage{microtype}      
\usepackage{lipsum}

\newcommand{\boma}[1]{\mbox{\boldmath ${#1}$}}

\title{Compositional Cubes: A new concept for multi-factorial compositions}

\author{
	Kamila~Fa\v cevicov\'a\\
	Department of Mathematical Analysis and Applications of Mathematics\\
	Palack\'y University Olomouc\\
	Czech Republic, 771 46 \\
	\texttt{kamila.facevicova@gmail.com} \\
	\And
	Peter~Filzmoser \\
	Institute of Statistics and Mathematical Methods in Economics\\
	Vienna University of Technology\\
	Austria, 1040\\
	\texttt{P.Filzmoser@tuwien.ac.at} \\
	\And
	Karel~Hron\\
	Department of Mathematical Analysis and Applications of Mathematics\\
	Palack\'y University Olomouc\\
	Czech Republic, 771 46 \\
	\texttt{karel.hron@upol.cz} \\
}

\begin{document}
\maketitle

\begin{abstract}
Compositional data are commonly known as multivariate observations carrying
relative information. Even though the case of vector or even two-factorial compositional data (compositional tables) is already well described in the literature, 
there is still a need for a comprehensive approach to the analysis of multi-factorial relative-valued data. Therefore, this contribution builds around the current knowledge about compositional 
data a general theory of work with $k$-factorial compositional data. 
As a main finding it turns out that similar to the case of compositional tables also the multi-factorial structures can be orthogonally decomposed into an independent and several interactive parts and, moreover, a coordinate representation allowing for their separate analysis by standard analytical methods can be constructed. For the sake of simplicity, these features are explained in detail for the case of three-factorial compositions (compositional cubes), followed by an outline
covering the general case. The three-dimensional structure is analysed in depth  in two practical examples, dealing with systems of spatial and time dependent compositional cubes. The methodology is implemented in the R package 
\texttt{robCompositions}.
\end{abstract}

\keywords{Analysis of independence \and Compositional data \and Coordinate representation \and Orthogonal decomposition}

\section{Introduction}
Consider a data set where the relative structure of parts is of the interest. As an example, the age structure of all employees in a given country can be analysed. In this case, ratios between parts rather than the absolute values, depending e.g. on the size of the country, are in the spotlight. As it will be shown later, this situation is already plentifully studied within the classical framework of compositional data analysis \citep{aitchison82,aitchison86}. On the other hand, when the structure is formed according to more than one factor, e.g. one can study the employment structure from the perspective of age and gender of employees, etc., the classical theory needs to be properly modified in order to respect such more complex structures. The first step towards this goal was dedicated to compositional tables, two factorial tables \citep{facevicova18}, but also structures formed by three or even more factors are likely to occur in practice. For instance, in addition to gender and age, one could be 
interested in analysing the employment structure according to full-time and 
part-time employment. Therefore, the manuscript introduces a general framework of dealing with multi-factorial compositional data, which extends the current state of knowledge acquired by compositional tables.

Processing of data which carry relative information is the goal of compositional data analysis that developed into a concise methodology with a wide 
range of possible applications, see, e.g., \cite{pawlowsky15}, \cite{filzmoser18} and references therein. 
A $D$-part composition is defined as a vector with positive components (parts) $\mathbf{x}=(x_1, \dots, x_D)'$, where the real information content is in the 
ratios between these parts rather than directly in the measured absolute values. 
In other words, compositional data describe quantitatively relative contributions of parts 
on a whole. Consequently, compositional data are scale invariant and can be represented 
without any loss of information as observations with a prescribed sum of the parts, 
e.g. in proportions (sum 1) or percentages (sum 100). Accordingly, the sample space of 
(representations of) compositional data is traditionally considered to be the $D$-part 
simplex $\mathcal{S}^D=\left\{\mathbf{x}=\left(x_1, x_2, \dots, x_D\right)'\mid x_i>0, \forall i, \sum_i{x_i}=\kappa\right\}$. Note that the constant $\kappa>0$, representing the sum of 
the compositional parts, can be chosen arbitrarily and it reduces the 
dimensionality of the sample space to $D-1$. Specific features of compositional data, particularly the scale invariance property, are captured by the Aitchison geometry \citep{pawlowsky01,billheimer01} with Euclidean vector space properties, defined for compositions $\mathbf{x},\mathbf{y}\in\mathcal{S}^D$ and a real constant $\alpha$ by operations of perturbation, powering and the Aitchison inner product,
\begin{equation}
	\label{ag}
	\begin{split}
		\mathbf{x}\oplus \mathbf{y} = (x_1y_1,\ldots ,x_Dy_D)', \
		\alpha \odot \mathbf{x} = (x_1^{\alpha},\ldots ,x_D^{\alpha})',\\
		\langle \mathbf{x},\mathbf{y}\rangle _A=\frac{1}{2D}\sum_{i=1}^D \sum_{j=1}^D 
		\ln \frac{x_i}{x_j} \ln \frac{y_i}{y_j},
	\end{split}
\end{equation}
respectively. 
As a consequence, a direct application of traditional multivariate statistical methods 
that rely on the Euclidean geometry in the real space \citep{eaton83} is not appropriate. 
Even though it would be possible to adapt them to the Aitchison geometry, it is more sensible to find a way how to express compositional data isometrically in the 
$(D-1)$-dimensional real space and proceed there, just by taking into account the 
specific interpretation of the new variables. In the compositional data analysis 
context this refers to isometric log-ratio (ilr) coordinates \citep{egozcue03}, 
which are orthonormal with respect to the Aitchison geometry. The main idea is to find a 
system of $D-1$ orthonormal basis vectors $\mathbf{e}_i$ of $\mathcal{S}^D$, where the new coordinates 
$\mathbf{z}=(z_1,\ldots,z_{D-1})'\in\mathbb{R}^{D-1}$ are obtained as
\begin{equation}
	\label{eq:ilr}
	z_i = \langle \mathbf{x}, \mathbf{e}_i\rangle_A, \quad \mathrm{for} \quad i=1, \dots, D-1.
\end{equation}

Since there does not exist a standard basis on $\mathcal{S}^D$, an option is to
use such an ilr coordinate system which has an advantageous interpretation under the given problem setting. From the definition, any ilr coordinate is a 
log-contrast, i.e. a linear combination $\xi_1\ln x_1+\ldots+\xi_D\ln x_D$ with $\sum_{i=1}^D\xi_i=0$. One popular approach for the construction of orthonormal 
coordinates was defined in \cite{egozcue05}. The aim is to construct a sequence of binary partitions of groups of compositional parts in order to obtain coordinates 
that are interpretable in terms of balances between these groups of parts. 
Accordingly, sequential binary partitions (SBP) are based on a systematical splitting of 
the compositional vector into two non-overlapping subcompositions, and the generating 
process ends after $D-1$ steps when each subcomposition is formed by only one part. 
The $i$-th step of the partition produces one vector of log-contrast coefficients 
$\boma{\xi}_i=(\xi_{i1}, \dots, \xi_{iD})'$ with $r_i$ parts 
$\xi_{i+}=\sqrt{\frac{s_i}{r_i(r_i+s_i)}}$ at the positions corresponding to parts 
from the first subcomposition formed by this step (denoted with $+$), 
$s_i$ parts  $\xi_{i-}=-\sqrt{\frac{r_i}{s_i(r_i+s_i)}}$ at the positions related to 
parts from the second subcomposition (denoted with $-$), and $0$ elsewhere. 
The coefficients are closely linked to basis vectors through the relation $\mathbf{e}_i=\exp(\boma{\xi}_i)$, and the resulting coordinates can be obtained from 
Equation~(\ref{eq:ilr}) or, without the need of enumeration of $\mathbf{e}_i$, directly as the log-contrast
\begin{equation}
	\label{eq:balances0}
	z_i=\sum_{j=1}^D\xi_{ij}\ln x_j, \quad \mathrm{for} \quad i=1, \dots, D-1,
\end{equation}
or
\begin{equation}
	\mathbf{z} = \mathbf{V}\ln(\mathbf{x}),
	\label{eq:matrix_log_contrast}
\end{equation}
where the contrast matrix $\mathbf{V}$ of type $(D-1)\times D$ has rows formed by $\boma{\xi}_i$. In any case, the ilr coordinates (balances) respect the following formula
\begin{equation}
	\label{eq:balances}
	z_i = \sqrt{\frac{r_is_i}{r_i+s_i}}\ln\frac{g(x_{i_1}, \dots, x_{i_{r_i}})}{g(x_{i'_1}, \dots, x_{i'_{s_i}})}, \quad \mathrm{for} \quad i=1, \dots, D-1,
\end{equation}
where $g(.)$ stands for the geometric mean, and values 
$i_1, \dots, i_{r_i}$ and $i'_1, \dots, i'_{s_i}$ label parts from the 
first $(+)$ and the second $(-)$ subcomposition, respectively. 
From Equation~(\ref{eq:balances}) it results that SBP produces coordinates in form of log-ratios between mean representations of two groups of parts, what gave them their 
name -- balances. Particularly in the case of vector compositional data, balances allow for a simple and natural interpretation and are frequently used in applications \citep{pawlowsky15}.

Although balances form a flexible class of orthonormal coordinates for vector compositional data, a further challenge is to develop a coordinate representation for the case when 
the whole is distributed according to two or more factors. The case of two-factorial compositional data \citep{egozcue08,egozcue15}, referred to as compositional tables, was recently intensively studied in \cite{facevicova14,facevicova16,facevicova18}, and a general coordinate representation was derived in the way so that it reflects the possibility of a decomposition of compositional tables into independent and interactive parts. Accordingly, the resulting ilr coordinates 
have the form of balances (independent part) and log-odds ratios (interactive part) and respect the dimensionality of the decomposed parts.

More specifically, compositional tables model a situation, 
when the relative structure of the data is determined by two factors. 
Accordingly, not only the relations within each factor, but also relations 
between them need to be analysed. As an example consider an employment structure in a given 
country, distributed according to the age of employees and their gender. 
Three types of questions arise: Is the proportion of females among the employees 
comparable to the proportion of males? Does any of the age groups outbalance? Does the age structure 
of employees depend on their gender? The first two questions focus exclusively on one factor, suppressing the effect of the other one. The last question, on the other hand, links information from both factors together. When the within-factor structure is analysed, 
the effect of the other factor can be suppressed by averaging across all its levels. 
This results in a standard compositional vector, and balances are then a natural way of its coordinate representation. Particularly, consider a table formed by two factors,
a row factor with $I$ levels, and a column factor with $J$ levels.
The whole information about the relations among the $I$ levels of the row factor is 
preserved in $I-1$ coordinates of the form
\begin{equation}
	\label{eq:tab_row_balance}
	z_i^\mathrm{r}=\sqrt{\frac{s_it_iJ}{s_i+t_i}}\ln\frac{\left[g(\mathbf{x}_{i_1\bullet})\cdots g(\mathbf{x}_{i_{s_i}\bullet})\right]^{1/{s_i}}}{\left[g(\mathbf{x}_{i'_1\bullet})\cdots g(\mathbf{x}_{i'_{t_i}\bullet})\right]^{1/{t_i}}}, \quad \mathrm{for} \quad i=1, 2, \dots, I-1,
\end{equation}
where $s_i$ and $t_i$ correspond to the respective step of the SBP performed on levels 
of the row factor. 
The indices $(i_1\bullet, \ldots ,i_{s_i}\bullet)$ and 
$(i'_1\bullet,\ldots ,i'_{t_i}\bullet)$
specify the rows, and $g(\cdot)$ is the geometric mean.
Similarly, relations among $J$ levels of the column factor are preserved in
$J-1$ balances
\begin{equation}
	\label{eq:tab_column_balance}
	z_j^\mathrm{c}=\sqrt{\frac{u_jv_jI}{u_j+v_j}}\ln\frac{\left[g(\mathbf{x}_{\bullet j_1})\cdots g(\mathbf{x}_{\bullet j_{u_j}})\right]^{1/{u_j}}}{\left[g(\mathbf{x}_{\bullet j'_1})\cdots g(\mathbf{x}_{\bullet j'_{v_j}})\right]^{1/{v_j}}}, \quad \mathrm{for} \quad j=1, 2, \dots, J-1,
\end{equation}
constructed with respect to the SBP of levels of the column factor, where 
indices $(\bullet j_1,\ldots ,\bullet j_{u_j})$ and 
$(\bullet j'_1, \ldots , \bullet j'_{v_j})$ specify the included columns.

The relations between two factors are traditionally described by odds ratios \citep{agresti02}. This concept can be adapted to compositional tables, because the last group of coordinates has the form of log-odds ratios between four groups of parts. 
These groups are uniquely defined by row and column SBPs and represented by geometrical means of their parts. More specifically, these $(I-1)(J-1)$ odds ratio coordinates are given by
\begin{equation}
	\label{eq:tab_odds_ratio}
	z_{ij}^\mathrm{OR}=\sqrt{\frac{\mid A_{ij}\mid\mid D_{ij}\mid}{\mid A_{ij}\mid+\mid B_{ij}\mid +\mid C_{ij}\mid +\mid D_{ij}\mid }}\ln\frac{g(\mathbf{x}_{A_{ij}})g(\mathbf{x}_{D_{ij}})}{g(\mathbf{x}_{B_{ij}})g(\mathbf{x}_{C_{ij}})}
\end{equation}
for $i=1,2,\dots, I-1$ and $j=1,2,\dots, J-1$,
where $A_{ij}, \dots, D_{ij}$ are indices of parts in each group defined by the $i$-th and $j$-th step of row and column SBP, respectively, and  $\mid A_{ij}\mid, \dots, \mid D_{ij}\mid$ are the 
numbers of parts within these groups. The construction and interpretation of this coordinate system is discussed in detail in \cite{facevicova18}.

An important feature of compositional tables is the possibility of their 
orthogonal decomposition. In the ideal situation, where there exists no relationship 
between row and column factors, all parts of the compositional table would be formed 
by the product of row and column marginals. This leads to so called 
independence table. The orthogonal complement to the independence table is called 
interaction table. Since the previously introduced coordinate system respects 
this decomposition, the independence table is characterized by row and column balances, 
and the interaction table by odds ratio coordinates, it is possible to analyse each part separately.

Even though the structure of compositional tables is 
already well described in the literature, a comprehensive approach for the analysis 
of multi-factorial relative-valued data is still lacking. 
Thus, the framework of compositional 
tables is extended to a general theory to work with $k$-factorial compositional data. Besides other findings, it turns out that also the multi-factorial structures can be decomposed orthogonally into an independent and several interactive parts and, moreover, a coordinate representation allowing for their separate analysis is provided. Although all 
considerations in the next section (Section \ref{sec:cubes}) are performed just for 
the case of three-factorial compositional data (called \textit{compositional cubes} in the following), they can be easily generalized to the case of more than three factors. 

The construction and interpretation of the proposed coordinate system is explained on an illustrative example in Section \ref{sec:example}, which also introduces the function implemented in the R package {\tt robCompositions} \citep{robCompositions}. The coordinates are used for the analysis of the employment structure of the European OECD countries, when the graphical comparison of countries as well as spatial clustering are provided. Moreover, the main sources of differences between the clusters are investigated by robust principal component analysis. Section \ref{sec:example2} introduces for an example of Austrian mobility data a strategy for the analysis of multi-factorial time series. The final Section 5 concludes.

\section{Compositional Cubes}
\label{sec:cubes}

In this section we simplify the main findings derived in \cite{facevicova18} for the
two-factorial situation, and consequently generalize them to the multi-factorial case.
Consider a relative structure formed according to three factors with $I$, $J$ and 
$K$ levels, respectively. Such a situation can be represented with a compositional cube and written in the form
\begin{equation}
	\textbf{x}=\left(\begin{array}{ccc|ccc|ccc} x_{111}&\cdots &x_{1J1}&&&&x_{11K}&\cdots &x_{1JK}\\ \vdots&\ddots&\vdots&&\cdots&&\vdots&\ddots&\vdots\\
		x_{I11}&\cdots&x_{IJ1}&&&&x_{I1K}&\cdots&x_{IJK}\end{array}\right),
\end{equation}
where $x_{ijk}>0, \forall i,j,k$, and vertical lines separate the 
levels of the third factor, called slices in the following. Since compositional cubes form a special case of the concept of $I\cdot J\cdot K$-part vector compositional data, all basic definitions can be accommodated for this case.

The sample space of compositional cubes is a subset of the $I\cdot J\cdot K$-part simplex
\begin{equation}
	\mathcal{S}^{IJK}=\left\{\mathbf{x}=(x_{111},\dots, x_{IJK})'\mid x_{ijk}>0,\quad \forall i,j,k;
	\sum_{i,j,k=1}^{I,J,K}x_{ijk}=\kappa \right\},
\end{equation}
representing the three-factorial structures with $I$ rows, $J$ columns and $K$ slices. The basic operations of the Aitchison geometry modify to 
\begin{equation}
	\mathbf{x}\oplus\mathbf{y}=(x_{ijk}\cdot y_{ijk})_{i,j,k=1}^{I,J,K} \quad \mathrm{and} \quad \alpha\odot\mathbf{x}=(x_{ijk}^{\alpha})_{i,j,k=1}^{I,J,K}.
\end{equation}

\subsection{Decomposition of Compositional Cubes}
\label{sec:decomposition}

Authors of \cite{egozcue08} proposed a decomposition of compositional tables into independent and
interactive parts (still compositional tables), which are mutually orthogonal, 
and whose perturbation again leads to the original compositional table. 
The independent part mimics independence of the factors. As in the standard case of  contingency tables, an assumption of independence means that the whole information about 
the relative structure of both factors is preserved in row and column marginals, and 
each entry of the table can be obtained as their product. In the compositional case, 
the only difference is that the arithmetic marginals are replaced by the geometric ones. 
When the factors are not independent, and the compositional table does not equal to the independence 
one, another table needs to be introduced. The interaction table, defined simply as a 
residual resulting from the difference between the original and the independence tables,
preserves the whole information about the relations between the factors and becomes mainly important when these relations are analysed. A similar idea can be utilized also 
in the case of compositional cubes, but due to the presence of pairwise and whole 
interactions, it is possible to further decompose the interactive part of the cube into additional four cubes, each preserving information about another source of association between the factors.

Similar to the case of compositional tables, also parts of the independence compositional 
cube are formed by the product of row, column and slice (geometric) marginals 
\begin{equation}
	x_{ijk}^\mathrm{ind}=g(\mathbf{x}_{i\bullet\bullet})g(\mathbf{x}_{\bullet j\bullet})g(\mathbf{x}_{\bullet\bullet k}),
\end{equation}
where dots in the index indicate an aggregation over the respective factors.
In case of perfect independence of all three factors, the original cube would be equal to the independent one. 
Otherwise, all associations between the factors are preserved in the interactive part
\begin{equation}
	\mathbf{x}^\mathrm{int}=\mathbf{x}\ominus\mathbf{x}^\mathrm{ind}.
\end{equation}

As mentioned above, the interactive part can be further decomposed. First, the 
relations between row and column factors are analyzed. Aggregation over values of 
the slice factor eliminates its impact and reduces the three-dimensional structure to 
a system of $K$ similar compositional tables, forming $K$ slices of a cube. 
According to \cite{egozcue08}, the interactive part of a table is extracted by 
a division of its parts by the respective geometric marginals. These considerations 
result in the compositional cube $\mathbf{int}^{\mathrm{rc}}(\mathbf{x})$ with cells
\begin{equation}\label{eq:int_rc}
	\mathrm{int}^\mathrm{rc}(\mathbf{x})_{ijk}=\frac{g(\mathbf{x}_{ij\bullet})}{g(\mathbf{x}_{i\bullet\bullet})g(\mathbf{x}_{\bullet j\bullet})}.
\end{equation}
From formula (\ref{eq:int_rc}) follows that the $\mathbf{int}^{\mathrm{rc}}(\mathbf{x})$ is actually formed by $K$ equal slices (compositional tables). Moreover, the row and column geometric marginals (it means compositional tables resulting from aggregation of cube cells by geometric means across the respective direction) are uniform, which underlines the favorable structure of the proposed decomposition. The system of marginals is completed in the direction of slices, whose respective marginal table 
corresponds to (\ref{eq:int_rc}).
In order to extract the pure interaction between row and slice factor, the effect of 
the column factor needs to be filtered out using the geometric mean, and similarly as $\mathbf{int}^{\mathrm{rc}}(\mathbf{x})$, the row-slice interaction cube $\mathbf{int}^{\mathrm{rs}}(\mathbf{x})$ has parts
\begin{equation}\label{eq:int_rs}
	\mathrm{int}^\mathrm{rs}(\mathbf{x})_{ijk}=\frac{g(\mathbf{x}_{i\bullet k})}{g(\mathbf{x}_{i\bullet\bullet})g(\mathbf{x}_{\bullet\bullet k})}.
\end{equation}
Similarly to the case of $\mathbf{int}^{\mathrm{rc}}(\mathbf{x})$, also this cube has uniform marginals. This property holds for the row and slice directions and the marginal table computed across the columns equal to (\ref{eq:int_rs}).
Finally, interactions between column and slice factors are contained in the cube $\mathbf{int}^{\mathrm{cs}}(\mathbf{x})$ with cells
\begin{equation}\label{eq:int_cs}
	\mathrm{int}^\mathrm{cs}(\mathbf{x})_{ijk}=\frac{g(\mathbf{x}_{\bullet jk})}{g(\mathbf{x}_{\bullet j\bullet})g(\mathbf{x}_{\bullet\bullet k})}.
\end{equation}
Since this cube is formed by $I$ identical rows, also row marginals equal to a table with parts (\ref{eq:int_cs}), however the column and slice marginals are again uniform, i.e., they are composed by the same positive elements.
All pairwise interaction cubes are orthogonal, but since there was always one factor omitted from the consideration, the information about the interactive part of the original cube 
is still not complete. The structure of a compositional cube is completed by considering mutual interactions between all three factors. This corresponds to the cube 
\begin{equation}\label{eq:int_rcs}
	\mathbf{int}^\mathrm{rcs}(\mathbf{x})=\mathbf{x}^\mathrm{int}\ominus\mathbf{int}^\mathrm{rc}(\mathbf{x})\ominus\mathbf{int}^\mathrm{rs}(\mathbf{x})\ominus\mathbf{int}^\mathrm{cs}(\mathbf{x})
\end{equation}
with parts
\begin{equation}
	\mathrm{int}^\mathrm{rcs}(\mathbf{x})_{ijk}=\frac{x_{ijk}g(\mathbf{x}_{i\bullet\bullet})g(\mathbf{x}_{\bullet j\bullet})g(\mathbf{x}_{\bullet\bullet k})}{g(\mathbf{x}_{ij\bullet})g(\mathbf{x}_{\bullet jk})g(\mathbf{x}_{i\bullet k})}.
\end{equation}
Also this cube has an advantageous structure from the perspective of the marginal tables, which are in this case uniform in all three directions. Note here that a similar property holds also for the interactive part of a compositional table, where row and column marginals are from their construction uniform.
Such a decomposition of the multi-factorial data can be very useful for an in-depth analysis of the data structure, as it
has been demonstrated in \cite{facevicova16,facevicova18, facevicova21} for compositional tables. 
This can be expected also for compositional cubes. In particular, from the decomposition follows that for those interactions, for which the respective cubes in the interaction part are constructed, the respective marginals are uniform, i.e., the information is fully captured by these cubes and does not propagate further. Consequently, vector (one-factorial) geometric marginals occur in the decomposition indeed only in the independent part, as expected, and more-dimensional nontrivial marginals (here in form of compositional tables) are left for bifactorial interaction cubes.

On the other hand, if each part of the decomposition is considered separately, two challenges need to be taken into account. 
At first, the dimensions of the sample spaces of the decomposed parts differ from the 
overall dimension $IJK -1$ of the original cube: for the independence cube $\mathbf{x}^{\mathrm{ind}}$ the dimension equals $I+J+K-3$, for cube $\mathbf{int}^{\mathrm{rc}}(\mathbf{x})$ it is $(I-1)(J-1)$ (and similarly for the remaining cubes related to paired interactions) and, finally, the sample space of cube $\mathbf{int}^\mathrm{rcs}(\mathbf{x})$ has dimension $(I-1)(J-1)(K-1)$. The reduced dimension can cause computational problems when an arbitrary ilr coordinate system (primarily designed for vector compositional data) is used for the representation of independence and interaction cubes. 
For example, this can be the case for robust statistical analysis \citep{rendlova18}, 
but also in general it is desirable to assign to each of the cubes from the decomposition 
such a number of coordinates (out of the total number $IJK -1$ of them) that reflects their respective dimensionality. The second problem concerns the interpretation of the results. 
Even though it is usually possible to convert the results back to the simplex, it is convenient to proceed with the analysis in some well-interpretable coordinates. 
Obviously, balances, defined as a log-ratio between two groups of parts, are not able to capture the multi-factorial nature of the compositional cubes. Although they can help to describe the relative structure of each factor separately, for a description of 
interactions we need to construct some alternative coordinate system. The construction of such orthonormal coordinates is presented in the following section. 

\subsection{Coordinate Representation of Compositional Cubes}
\label{sec:coord_rep}

In this section we will focus on a possible coordinate representation of three-factorial compositional data, compositional cubes, which simplifies substantially 
the construction of ilr coordinates for compositional tables proposed in \cite{facevicova18}. A deeper understanding of the structure of this coordinate representation allows its generalisation and application to compositional data describing relationships given by more than three factors. 
In order to keep the construction as simple as possible, we consider a vectorized version of the cube
\begin{equation}
	\mathrm{vec}(\mathbf{x}) = \left(x_{111}, x_{112}, \dots, x_{1JK}, x_{211}, \dots, x_{2JK}, \dots, x_{I11}, \dots, x_{IJK} \right)'.
\end{equation}

As it was already suggested, balances can help to describe the relative structure within each factor. For this purpose, the whole rows, columns and slices (each represented by geometric mean across all levels of the remaining factors) should be taken. After $I-1$ steps of the sequential binary partition applied on levels of the row factor (SBPr), a system of $I-1$ vectors $\boma{\xi}_i^{\mathrm{r}}$ (of length $IJK$) is obtained. The $i$-th generating vector has entries 
$\xi_{i+}^{\mathrm{r}} = \sqrt{\frac{t_i}{s_i(s_i+t_i)JK}}$ at positions corresponding to parts from the $s_i$ rows of the cube $\mathbf{x}$, which were in the respective step assigned to the $+$ group, and 
$\xi_{i-}^{\mathrm{r}} = -\sqrt{\frac{s_i}{t_i(s_i+t_i)JK}}$ at positions corresponding to parts from the $t_i$ rows assigned to the $-$ group, and zero elsewhere. The first group of coordinates is thus simply formed by row balances, which characterize the structure of the row factor when the influence of the other factors is suppressed,
\begin{equation}
	z_i^\mathrm{r}=\sqrt{\frac{s_it_iJK}{s_i+t_i}}\ln\frac{\left[g(\mathbf{x}_{i_1\bullet\bullet})\cdots g(\mathbf{x}_{i_{s_i}\bullet\bullet})\right]^{1/s_i}}
	{\left[g(\mathbf{x}_{i'_1\bullet\bullet})\cdots g(x_{i'_{t_i}\bullet\bullet})\right]^{1/t_i}}, \quad \mathrm{for} \quad i=1, 2, \dots, I-1.
\end{equation}
A similar construction can be made for column and slice factors. In the first case, 
a sequential binary partition of the whole columns (SBPc) results in a system of $J-1$ vectors $\boma{\xi}_j^{\mathrm{c}}$ with entries 
$\xi_{j+}^{\mathrm{c}} = \sqrt{\frac{v_j}{u_j(u_j+v_j)IK}}$, $\xi_{j-}^{\mathrm{c}} = -\sqrt{\frac{u_j}{v_j(u_j+v_j)IK}}$ and $0$, always corresponding to parts of 
$u_j$ columns from the $+$ group, $v_j$ columns from the $-$ group, and columns not included to the respective step of SBPc. The inner structure of the column factor is preserved through column balances
\begin{equation}
	z_j^\mathrm{c}=\sqrt{\frac{u_jv_jIK}{u_j+v_j}}\ln\frac{\left[g(\mathbf{x}_{\bullet j_1\bullet})\cdots g(\mathbf{x}_{\bullet j_{u_j}\bullet})\right]^{1/u_j}}
	{\left[g(\mathbf{x}_{\bullet j'_1\bullet})\cdots g(\mathbf{x}_{\bullet j'_{v_j}\bullet})\right]^{1/v_j}}, \quad \mathrm{for} \quad j=1, 2, \dots, J-1,
\end{equation}
which form the second group in the coordinate system representing the whole compositional cube $\mathbf{x}$ as well as its independent part $\mathbf{x}^{\mathrm{ind}}$. The third group describes the structure of the slice factor. A sequential binary partition of the whole slices (SBPs) now determines the final system of $K-1$ vectors $\boma{\xi}_k^{\mathrm{s}}$ with entries 
$\xi_{k+}^{\mathrm{s}} = \sqrt{\frac{n_k}{m_k(m_k+n_k)IJ}}$, $\xi_{k-}^{\mathrm{s}} = -\sqrt{\frac{m_k}{n_k(m_k+n_k)IJ}}$ and $0$, corresponding to $m_k$ slices from group $+$, $n_k$ slices from group $-$, and the remaining slices not included in the $k$-th step, respectively; the slice balances are
\begin{equation}
	z_k^\mathrm{s}=\sqrt{\frac{m_kn_kIJ}{m_k+n_k}}\ln\frac{\left[g(\mathbf{x}_{\bullet\bullet k_1})\cdots g(\mathbf{x}_{\bullet\bullet k_{m_k}})\right]^{1/m_k}}
	{\left[g(\mathbf{x}_{\bullet\bullet k'_1})\cdots g(\mathbf{x}_{\bullet\bullet k'_{n_k}})\right]^{1/n_k}}, \quad \mathrm{for} \quad k=1, 2, \dots, K-1.
\end{equation}

Note here that the row, column and slice balances form a complete coordinate representation of $\mathbf{x}^\mathrm{ind}$, since all the remaining coordinates of the independence cube are zero. 

When row, column and slice SBPs are defined, we can immediately construct the remaining elements of the coordinate system of cube $\mathbf{x}$, which also correspond to coordinates of $\mathbf{x}^\mathrm{int}$. For this purpose the normalized Hamadard (entry wise) product ($\circ$) of the vectors $\boma{\xi}_i^{\mathrm{r}}$, $\boma{\xi}_j^{\mathrm{c}}$ and $\boma{\xi}_k^{\mathrm{s}}$ and Equation~(\ref{eq:balances0}) is used. The vectors
$\boma{\xi}_{ij}^{\mathrm{rc}} = \boma{\xi}_i^{\mathrm{r}}\circ\boma{\xi}_j^{\mathrm{c}}$, $i=1,\dots, I-1$, $j=1, \dots, J-1$ determine $(I-1)(J-1)$ coordinates of type
\begin{equation}
	z_{ij}^\mathrm{rc}=\sqrt{\frac{\mid A_{ij}\mid \mid D_{ij}\mid }{\mid A_{ij}\mid +\mid B_{ij}\mid +\mid C_{ij}\mid +\mid D_{ij}\mid }}\ln\frac{g(\mathbf{x}_{A_{ij}})g(\mathbf{x}_{D_{ij}})}{g(\mathbf{x}_{B_{ij}})g(\mathbf{x}_{C_{ij}})}, 
\end{equation}
for $i=1, 2, \dots, I-1$ and $j=1, 2, \dots, J-1$,
capturing the interactions between row and column factors, which through the geometric mean suppress the influence of the slice factor. Obviously, these coordinates are formed by four groups of parts 
(denoted as $A, B, C, D$ and represented by their respective geometric means), 
and can be interpreted in terms of a log-odds ratio, which is also 
used in a standard statistical analysis of two-factorial data \citep{agresti02}. 
This system of coordinates, composed into the ($IJK-1$)-component vector 
$\mathbf{z}^{\mathrm{rc}}$ with $z_{ij}^\mathrm{rc}$ on $(I-1)(J-1)$ positions 
corresponding to $\mathbf{int}^\mathrm{rc}(\mathbf{x})$ and zeros elsewhere, thus 
allows for the analysis of the relations exclusively between row and column factors.

The Hadamard product of $\boma{\xi}_i^{\mathrm{r}}$ and $\boma{\xi}_k^{\mathrm{s}}$, $\boma{\xi}_{ik}^{\mathrm{rs}}$, $i=1,\dots, I-1$, $k=1, \dots, K-1$, leads to coordinates
\begin{equation}
	z_{ik}^\mathrm{rs}=\sqrt{\frac{\mid A'_{ik}\mid \mid D'_{ik}\mid }{\mid A'_{ik}\mid +\mid B'_{ik}\mid +\mid C'_{ik}\mid +\mid D'_{ik}\mid }}\ln\frac{g(\mathbf{x}_{A'_{ik}})g(\mathbf{x}_{D'_{ik}})}{g(\mathbf{x}_{B'_{ik}})g(\mathbf{x}_{C'_{ik}})},
\end{equation}
for $i=1, 2, \dots, I-1$ and $k=1, 2, \dots, K-1$,
which capture the information about the relations between row and slice factors, when the 
influence of the column factor is suppressed. 
Similar to the case of $\mathbf{z}^{\mathrm{rc}}$, also coordinates contained in the respective vector $\mathbf{z}^{\mathrm{rs}}$ can be interpreted in terms of 
a log-odds ratio and are utilized, when the relationship between row and slice factors 
is of primary interest.

Finally, the Hadamard products of $\boma{\xi}_j^{\mathrm{c}}$ and $\boma{\xi}_k^{\mathrm{s}}$, $j=1,\dots, J-1$, $k=1, \dots, K-1$ lead to vectors $\boma{\xi}_{jk}^{\mathrm{cs}}$ and coordinates
\begin{equation}
	z_{jk}^\mathrm{cs}=\sqrt{\frac{\mid A''_{jk}\mid \mid D''_{jk}\mid }{\mid A''_{jk}\mid +\mid B''_{jk}\mid +\mid C''_{jk}\mid +\mid D''_{jk}\mid }}\ln\frac{g(\mathbf{x}_{A''_{jk}})g(\mathbf{x}_{D''_{jk}})}{g(\mathbf{x}_{B''_{jk}})g(\mathbf{x}_{C''_{jk}})}, 
\end{equation}
for $j=1, 2, \dots, J-1$ and $k=1, 2, \dots, K-1$.
The system of these coordinates $\mathbf{z}^{\mathrm{cs}}$ completes the odds ratio-type coordinates with those concerning relations between column and slice factors.

To complete the original data structure, also full interactions between all three factors 
need to be contained in the coordinate system. The remaining $(I-1)(J-1)(K-1)$ coordinates 
are determined by the Hadamard product of all three types of vectors, $\boma{\xi}_i^{\mathrm{r}}, \boma{\xi}_j^{\mathrm{c}}$ and $\boma{\xi}_k^{\mathrm{s}}$, $i=1, \dots, I-1$, $j=1,\dots, J-1$, $k=1, \dots, K-1$, and have a general form
\begin{equation}
	z_{ijk}^\mathrm{rcs}=Q_{ijk}\ln\frac{g(\mathbf{x}_{A'''_{ijk}})g(\mathbf{x}_{D'''_{ijk}})g(\mathbf{x}_{F'''_{ijk}})g(\mathbf{x}_{G'''_{ijk}})}{g(\mathbf{x}_{B'''_{ijk}})g(\mathbf{x}_{C'''_{ijk}})g(\mathbf{x}_{E'''_{ijk}})g(\mathbf{x}_{H'''_{ijk}})}, 
\end{equation}
for $i=1, 2, \dots, I-1$, $j=1, 2, \dots, J-1$, and $k=1, 2, \dots, K-1$,
where 
{\footnotesize
	\begin{equation}
		Q_{ijk}=\sqrt{\frac{\mid A'''_{ijk}\mid \mid H'''_{ijk}\mid }{\mid A'''_{ijk}\mid +\mid B'''_{ijk}\mid +\mid C'''_{ijk}\mid +\mid D'''_{ijk}\mid +\mid E'''_{ijk}\mid +\mid F'''_{ijk}\mid +\mid G'''_{ijk}\mid +\mid H'''_{ijk}\mid }},
	\end{equation}	
}
is a constant ensuring orthonormality of the coordinates.
Even though the interpretation of this last group of coordinates may be a bit tricky (one possible interpretation is in terms of a log-ratio of two odds ratios), their definition is necessary to complete the system of $IJK-1$ orthonormal coordinates of the original table $\mathbf{x}$. Moreover, when a sample of compositional cubes is available, these coordinates can be 
used for instance to test for the presence of full interactions. 

\begin{figure}
	\centering
	\includegraphics[width=0.7\linewidth]{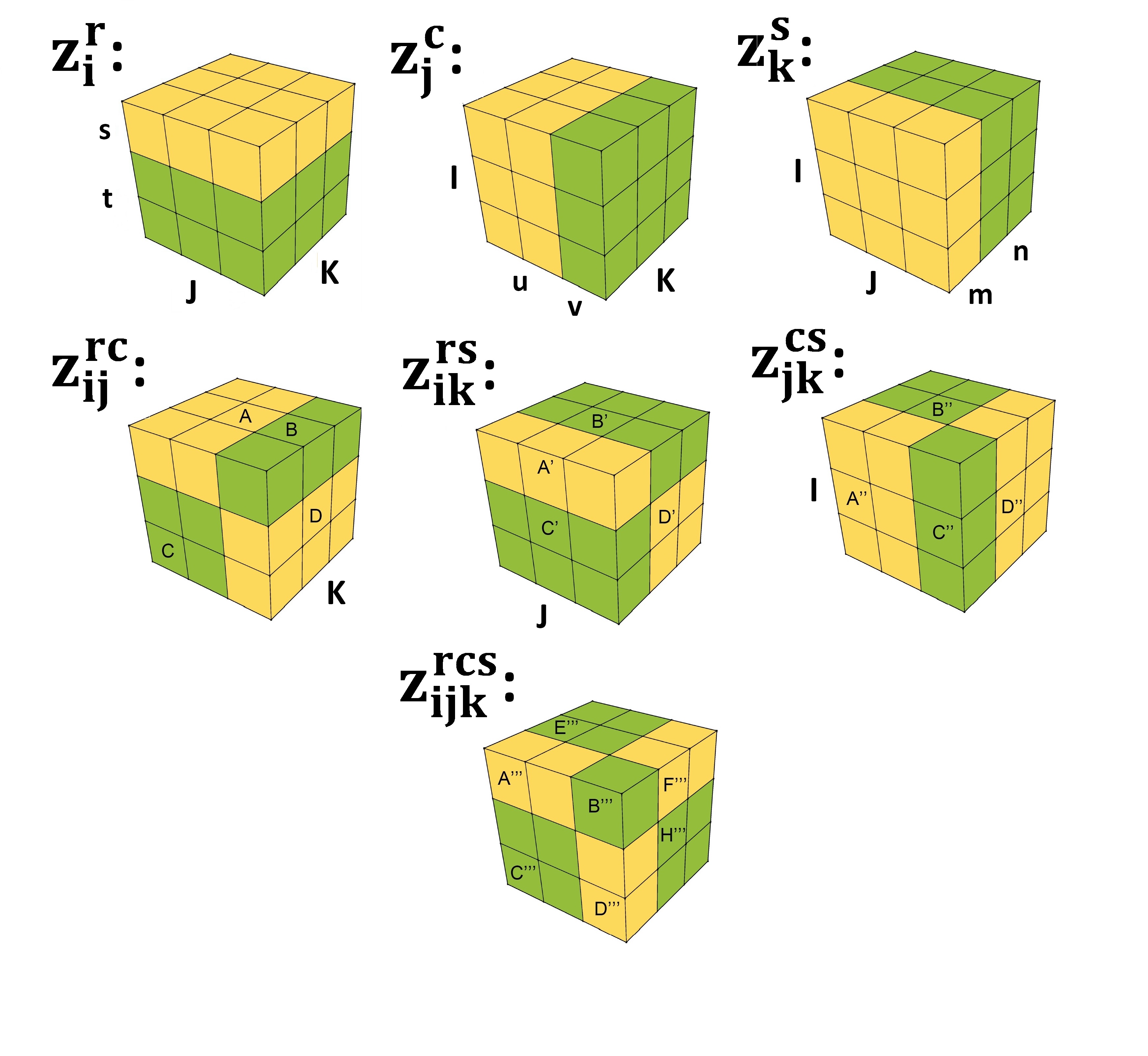}
	\caption{Graphical representation of groups of parts involved in each type of coordinates forming the whole coordinate system designed for compositional cubes.}
	\label{fig:krychle_deleni}
\end{figure}

For an easier understanding of the coordinate structure, especially the assignment of 
parts into groups, Figure~\ref{fig:krychle_deleni} provides a graphical representation of each type of proposed coordinates. The specific interpretation of the coordinates will be explained on a practical example in Section~\ref{sec:example}.

Besides the benefits of the proposed coordinate system in terms of interpretation, 
it is important to point out that the coordinates reflect the dimensionality of the 
sample space of the decomposed parts to which they are assigned, and thus allow to analyse these parts separately. Of course, each decomposed part is still a cube of the same dimension as the original one (with $I$ rows, $J$ columns and $K$ slices) and its coordinate representation must contain $IJK-1$ components, but the structure of the vector of coordinates follows the one introduced for $\mathbf{z}^{\mathrm{rc}}$ of the interaction cube $\mathbf{int}^\mathrm{rc}(\mathbf{x})$.
More specifically, e.g.~the cube $\mathbf{int}^\mathrm{rc}(\mathbf{x})$ 
is in the proposed system represented with coordinates $z_{ij}^\mathrm{rc}$ and a vector of zeros resulting from application of the remaining coordinate formulas applied on this partial cube. 
Accordingly, with respect to the decomposition described in Section~\ref{sec:decomposition}, 
for the coordinate representation $\mathbf{z}$ of the original compositional cube the following relation
holds,
\begin{equation}
	\label{eq:coord_cubes}
	\mathbf{z}=\mathbf{z}^{\mathrm{r}}+\mathbf{z}^{\mathrm{c}}+\mathbf{z}^{\mathrm{s}}+\mathbf{z}^{\mathrm{rc}}+\mathbf{z}^{\mathrm{rs}}+\mathbf{z}^{\mathrm{cs}}
	+\mathbf{z}^{\mathrm{rcs}}.
\end{equation}

Even though the interpretation of the coordinate system is determined by the initial SBPs, any other relationship within the compositional cube is reachable through a transformation matrix $\mathbf{T}$, whose rows are formed by coefficients of the respective logarithmized parts of $\mathrm{vec}(\mathbf{x})$ in the desirable log-ratios. According to (\ref{eq:matrix_log_contrast}), the vectorized form of a compositional cube $\mathrm{vec}(\mathbf{x})$ is equal (after closure) to $\exp(\mathbf{V}'\mathbf{z})$. Therefore, a system of log-contrasts representing a given compositional cube equals
\begin{equation}
	\mathbf{z}^\ast=\mathbf{T}\mathbf{V}'\mathbf{z}.
	\label{eq:lib_log_ratio}
\end{equation}

\subsection{General Properties of Multi-factorial Compositional Data}

The findings from Sections \ref{sec:decomposition} and \ref{sec:coord_rep} can be directly extended to a general $k$-factorial case. $k$-factorial compositional data are formed by a $k$-dimensional array of positive entries, representing a relative structure given by levels of $k$ constituting factors. Also such a complex structure contains its independent and interactive parts, where the independent part equals to the product of (vector) geometrical marginals. The sources of interactions are given by relations between pairs, triplets, quaternions, etc. of constituting factors and therefore the interactive part can be further orthogonally decomposed to objects carrying information about each of these sources. The main principle is based on aggregation over the redundant dimensions and expression of interactions within the resultant object. In the case of compositional cubes we have seen that pairwise interactions actually correspond to the interactive part of a compositional table formed by geometric means computed across levels of the third factor, see e.g. Equation (\ref{eq:int_rc}). Similarly, in the case of a four-factorial compositional object, all sources of interactions between a selected triplet of factors can be reached by a decomposition of a cube given by aggregation over the remaining fourth dimension. When we vary over the fourth dimension, all pairwise and three-way interactions are extracted and, finally, by subtraction of all these parts together with the independent one, the object preserving the full interactions is reached (similarly as in the Equation (\ref{eq:int_rcs})).

Section \ref{sec:coord_rep} shows that the whole coordinate representation of a compositional cube is determined by three systems of SBPs,
separately given for the levels of row, column and slice factors. Similarly, also $k$-factorial compositions can be represented in orthonormal coordinates. Balances between levels of the individual factors characterize the independent part of the object. Log-constrasts obtained from the Hadamard product of pairs, triplets, etc. of SBP basis vectors $\boma{\xi}$ and Equation~(\ref{eq:balances0}) then represent the respective sources of interactions.

\section{Example: Employment Structure}
\label{sec:example}

\begin{figure}[htp]
	\centering
	\includegraphics[width=0.3\linewidth]{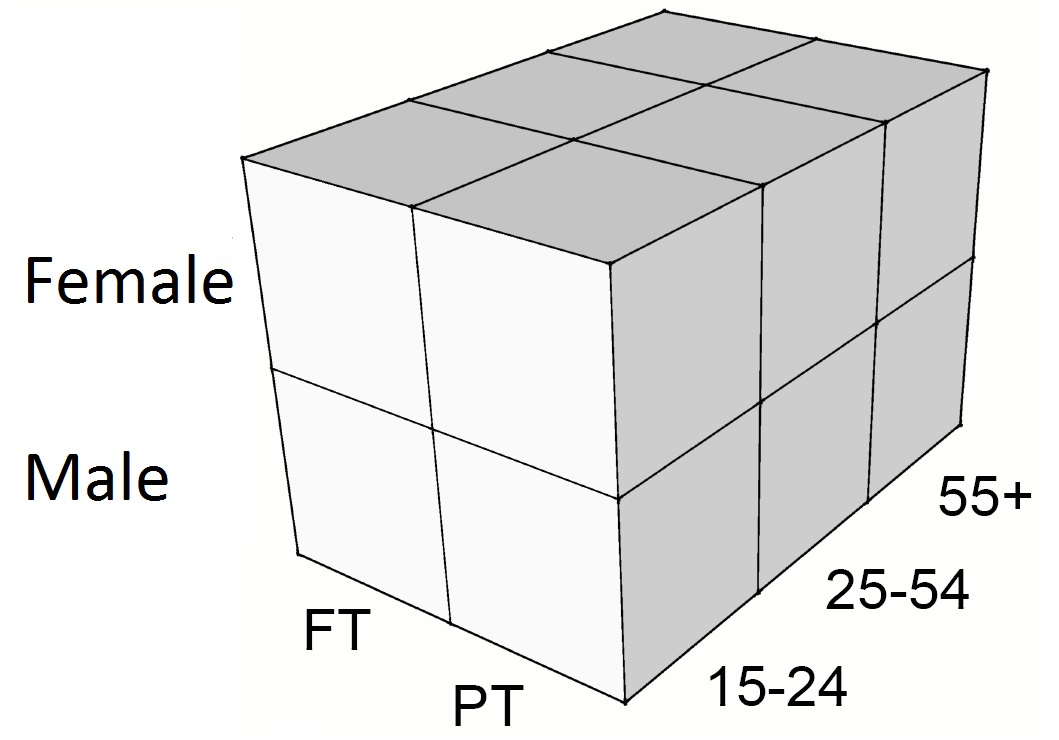}
	\caption{Graphical representation of the cube structure used in Section \ref{sec:example}. Rows represent gender of employees, columns type of contract (FT: full-time, PT: part-time) and slices separate different age groups.}
	\label{fig:cube}
\end{figure}

The use of the proposed approach can be demonstrated on an example 
where the analysis of the employment structure in several countries is of interest. 
For this purpose, data from 32 European members of OECD were collected at \url{http://stats.oecd.org}. For each country in the sample, an estimated number of employees 
in the year 2015 was available. The data were structured according to gender and age of employees and 
the type of their contract. More specifically, we distinguish males (M) and females (F), 
young (category 15--24), middle-aged (25--54) and older (55+) employees, and full-time (FT) and part-time (PT) contracts. The data at hand thus form a sample of 32 cubes with two 
rows (gender), two columns (type of contract) and three slices (age), which allow for a 
deeper analysis of the overall employment structure, not just from the perspective of 
each factor separately, but also from the perspective of the relations/interactions 
between them. Besides the global aspects of the employment, the analysis aims also 
at revealing the national specifics of the countries contained in the sample. An example 
of one cube from Czech Republic is displayed in Table 1, and a graphical 
overview of the cubes is depicted in Figure \ref{fig:cube}.

\begin{table}[htp]
	\begin{center}
		\begin{minipage}{\textwidth}
			\caption{Example of one cube from the sample analyzed in Section \ref{sec:example}: employment structure in the Czech Republic in 2015 (in thousands of employees).}
			\label{tab:CR}
			\begin{tabular*}{\textwidth}{@{\extracolsep{\fill}}lcccccc@{\extracolsep{\fill}}}
				\toprule
				Gender&\multicolumn{2}{c}{$15 - 24$}&\multicolumn{2}{c}{$25 - 54$}&\multicolumn{2}{c}{$55+$}\\
				&FT&PT&FT&PT&FT&PT\\
				\midrule
				Female&104.756&17.128&1618.415&90.505&317.031&56.355\\
				Male&169.851&11.165&2127.849&22.759&467.212&38.208\\
				\bottomrule
			\end{tabular*}
		\end{minipage}
	\end{center}
\end{table}

Obviously, the counts in the cells of the cubes depend on the population 
size of the country. On the other hand, when the analysis of structural patterns of the employment in several countries is of the interest, 
the compositional approach seems to be appropriate, because the population size 
is not be relevant in this approach. The analysis was performed using 
function {\tt cubeCoordWrapper} from the R package {\tt robCompositions} \citep{robCompositions}. 

\subsection{Coordinate Representation}

Following Section \ref{sec:cubes}, the row, column and slice SBPs need to be determined 
prior to the construction of coordinates. In case of age and gender factors, 
only their two levels need to be separated. Consequently, the first two generating vectors 
are
$$\boma{\xi}^\mathrm{r}_1=\sqrt{\frac{1}{12}}(1,1,1,1,1,1,-1,-1,-1,-1,-1,-1)'$$ 
and 
$$\boma{\xi}^\mathrm{c}_1=\sqrt{\frac{1}{12}}(1,1,1,-1,-1,-1,1,1,1,-1,-1,-1)',$$
where the components of these vectors correspond to the cells of the vectorized form of the 
cube, 
\begin{displaymath}
	\label{eq:ex_vectorized}	\mathrm{vec}(\mathbf{x})=(x_{111}, x_{112}, x_{113}, x_{121}, x_{122}, x_{123}, x_{211}, x_{212}, x_{213}, x_{221}, x_{222}, x_{223})'.
\end{displaymath}

There are more options of the slice SBP, where the analyst can decide which age 
group has to be separated first. Here, the youngest group was firstly separated from the 
remaining two groups and, in the next step, the middle-aged group (25--54 years) 
from the oldest. The other options, starting with the separation of the middle-aged or the oldest group, respectively, would lead to similar results (in terms of presence of interaction between factors), but they would slightly 
alter the interpretation. In the presented case, the generating vectors are
$$\boma{\xi}^\mathrm{s}_1=\sqrt{\frac{1}{6}}(1,-0.5, -0.5, 1, -0.5, -0.5, 1, -0.5, -0.5, 1, -0.5, -0.5)'$$ and $$\boma{\xi}^\mathrm{s}_2=\sqrt{\frac{1}{8}}(0,1,-1,0,1,-1,0,1,-1,0,1,-1)'.$$ 
Following the construction from Section \ref{sec:cubes}, the Hadamard product of the above derived generating vectors leads (after their normalization) to the remaining system of vectors. Particularly,

\begin{eqnarray}
	\nonumber
	\boma{\xi}^\mathrm{rc}_{11}&=&\sqrt{\frac{1}{12}}(1,1,1,-1,-1,-1,-1,-1,-1,1,1,1)'\propto \boma{\xi}^\mathrm{r}_1 \circ \boma{\xi}^\mathrm{c}_1, \\
	\nonumber
	\boma{\xi}^\mathrm{rs}_{11}&=&\sqrt{\frac{1}{6}}(1,-0.5,-0.5,1,-0.5,-0.5,-1,0.5,0.5,-1,0.5,0.5)'\propto \boma{\xi}^\mathrm{r}_1 \circ \boma{\xi}^\mathrm{s}_1,\\
	\nonumber
	\boma{\xi}^\mathrm{rs}_{12}&=&\sqrt{\frac{1}{8}}(0,1,-1,0,1,-1,0,-1,1,0,-1,1)'\propto \boma{\xi}^\mathrm{r}_1 \circ \boma{\xi}^\mathrm{s}_2,\\
	\nonumber
	\boma{\xi}^\mathrm{cs}_{11}&=&\sqrt{\frac{1}{6}}(1,-0.5,-0.5,-1,0.5,0.5,1,-0.5,-0.5,-1,0.5,0.5)'\propto \boma{\xi}^\mathrm{c}_1 \circ \boma{\xi}^\mathrm{s}_1,\\
	\nonumber
	\boma{\xi}^\mathrm{cs}_{12}&=&\sqrt{\frac{1}{8}}(0,1,-1,0,-1,1,0,1,-1,0,-1,1)'\propto \boma{\xi}^\mathrm{c}_1 \circ \boma{\xi}^\mathrm{s}_2,\\
	\nonumber
	\boma{\xi}^\mathrm{rcs}_{111}&=&\sqrt{\frac{1}{6}}(1,-0.5,-0.5,-1,0.5,0.5,-1,0.5,0.5,1,-0.5,-0.5)'\propto \boma{\xi}^\mathrm{r}_1 \circ \boma{\xi}^\mathrm{c}_1 \circ \boma{\xi}^\mathrm{s}_1, \\
	\nonumber
	\boma{\xi}^\mathrm{rcs}_{112}&=&\sqrt{\frac{1}{8}}(0,1,-1,0,-1,1,0,-1,1,0,1,-1)'\propto \boma{\xi}^\mathrm{r}_1 \circ \boma{\xi}^\mathrm{c}_1\circ \boma{\xi}^\mathrm{s}_2.
\end{eqnarray}

\begin{figure}
	\centering
	\includegraphics[width=0.65\linewidth]{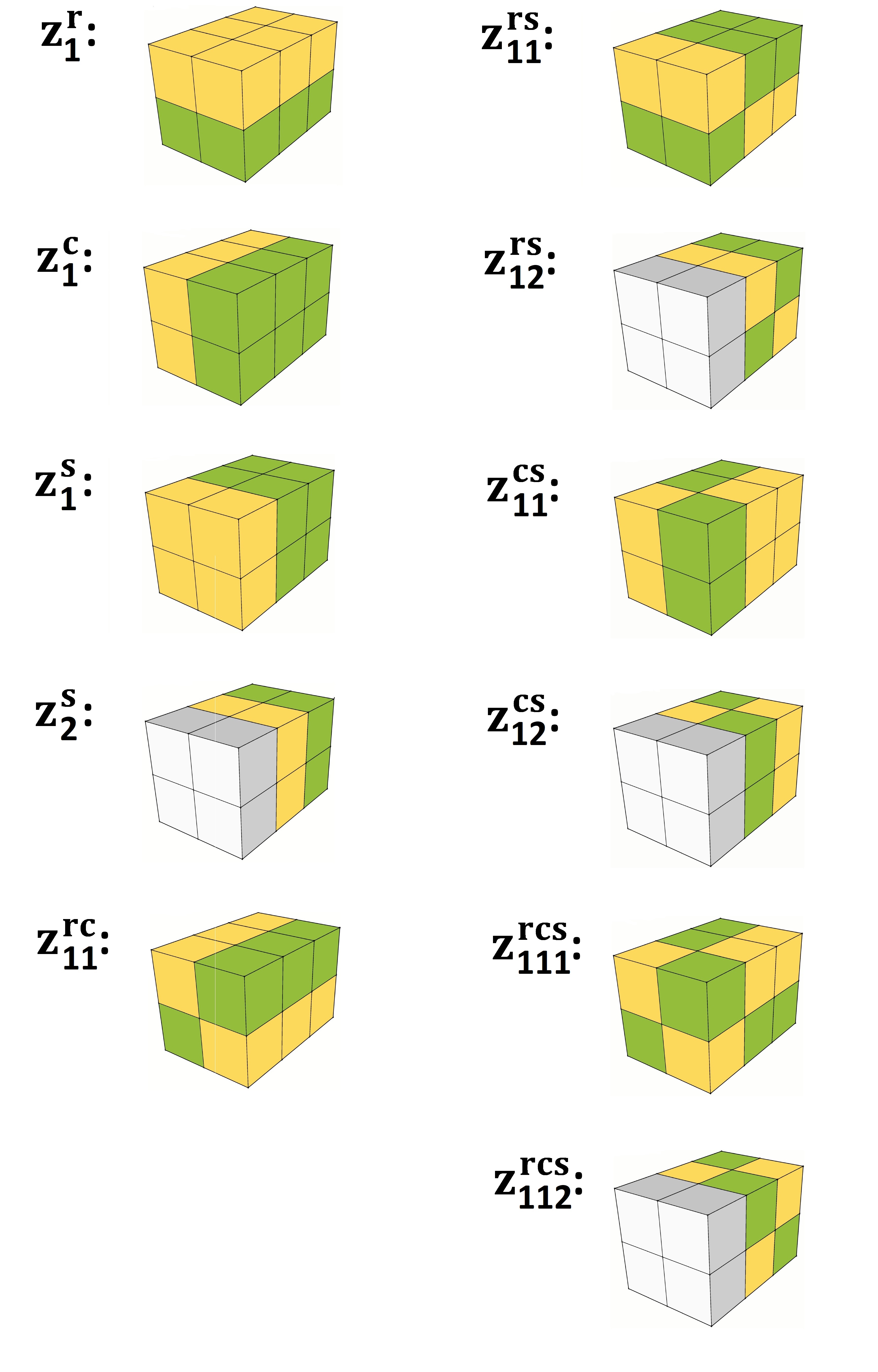}
	\caption{Graphical representation of the groups of parts involved in particular coordinates. Yellow parts constitute the numerator, green the denominator and white parts are not included in the respective log-ratio.}
	\label{fig:ex_coord}
\end{figure}

Finally, according to Equation (\ref{eq:balances0}), these vectors lead to a system of $11$ orthonormal coordinates:
\begin{displaymath}
	\begin{array}{rclrcl}
		z^\mathrm{r}_1&=&\sqrt{3}\ln\frac{g(\mathbf{x}_{1\bullet\bullet})}{g(\mathbf{x}_{2\bullet\bullet})}&
		z^\mathrm{rs}_{11}&=&\sqrt{\frac{2}{3}}\ln\frac{g(\mathbf{x}_{1\bullet 1})\sqrt{g(\mathbf{x}_{2\bullet2})g(\mathbf{x}_{2\bullet3})}}{g(\mathbf{x}_{2\bullet 1})\sqrt{g(\mathbf{x}_{1\bullet2})g(\mathbf{x}_{1\bullet3})}}\\
		z^\mathrm{c}_1&=&\sqrt{3}\ln\frac{g(\mathbf{x}_{\bullet1\bullet})}{g(\mathbf{x}_{\bullet2\bullet})}&
		z^\mathrm{rs}_{12}&=&\sqrt{\frac{1}{2}}\ln\frac{g(\mathbf{x}_{1\bullet2})g(\mathbf{x}_{2\bullet3})}{g(\mathbf{x}_{2\bullet2})g(\mathbf{x}_{1\bullet3})}\\
		z^\mathrm{s}_1&=&\sqrt{\frac{8}{3}}\ln\frac{g(\mathbf{x}_{\bullet\bullet1})}{\sqrt{g(\mathbf{x}_{\bullet\bullet2})g(\mathbf{x}_{\bullet\bullet3})}}&
		z^\mathrm{cs}_{11}&=&\sqrt{\frac{2}{3}}\ln\frac{g(\mathbf{x}_{\bullet 11})\sqrt{g(\mathbf{x}_{\bullet22})g(\mathbf{x}_{\bullet23})}}{g(\mathbf{x}_{\bullet 21})\sqrt{g(\mathbf{x}_{\bullet12})g(\mathbf{x}_{\bullet13})}}\\
		z^\mathrm{s}_2&=&\sqrt{2}\ln\frac{g(\mathbf{x}_{\bullet\bullet2})}{g(\mathbf{x}_{\bullet\bullet3})}&
		z^\mathrm{cs}_{12}&=&\sqrt{\frac{1}{2}}\ln\frac{g(\mathbf{x}_{\bullet 12})g(\mathbf{x}_{\bullet23})}{g(\mathbf{x}_{\bullet 13})g(\mathbf{x}_{\bullet22})}\\
		z^\mathrm{rc}_{11}&=&\sqrt{\frac{3}{4}}\ln\frac{g(\mathbf{x}_{11\bullet})g(\mathbf{x}_{22\bullet})}{g(\mathbf{x}_{12\bullet})g(\mathbf{x}_{21\bullet})}&
		z^\mathrm{rcs}_{111}&=&\sqrt{\frac{1}{6}}\ln\frac{x_{111}x_{221}\sqrt{x_{122}x_{123}}\sqrt{x_{212}x_{213}}}{x_{211}x_{121}\sqrt{x_{112}x_{113}}\sqrt{x_{222}x_{223}}}\\
		&&&z^\mathrm{rcs}_{112}&=&\sqrt{\frac{1}{8}}\ln\frac{x_{112}x_{222}x_{123}x_{213}}{x_{122}x_{212}x_{113}x_{223}}
		
	\end{array}	
\end{displaymath}
A graphical representation of these coordinates is provided in Figure~\ref{fig:ex_coord}.

\subsection{Interpretation}

The interpretation of the coordinates can be discussed on the example of data 
from the Czech Republic, see Table~1. The set of row, column and slice balances (as a subvector of $\mathbf{z}$) corresponds to 
$$(z^\mathrm{r}_1, z^\mathrm{c}_1, z^\mathrm{s}_1, z^\mathrm{s}_2)'=(0.304, 4.672, -2.487, 1.097)'.$$
These numbers are interpretable, as usual for balances, in terms of a dominance of either 
the group of cells in the numerator (positive value) or denominator (negative value) of the respective log-ratio. 
Accordingly, in the Czech Republic
the proportion of female employees slightly dominates over proportion of males ($z_1^r$), and full-time contracts
clearly dominate over part-time contracts ($z_1^c$).
The slice balances contain information about the age structure of the employees. 
Due to the high negative value of coordinate $z^\mathrm{s}_1$ it can be concluded 
that the youngest employees are outbalanced by those from the middle age and older groups;
within these latter groups, employees aged between 25--54 years prevail (coordinate $z^\mathrm{s}_2$). More specifically, the ratio between female and male employees is 1.19 (without the normalizing constant and logarithm), full-time contracts prevail the part-time almost fifteen times, group of 25+ employees is about 4.6 times bigger than group of youngest ones and, finally, there is about twice more employees aged between 25--54 than the oldest ones (55+). Another possible interpretation is in terms of an average log-ratio between the given groups of employees across all combinations of the remaining factors. E.g. if the coordinate $z^\mathrm{s}_1$ is divided by its normalizing constant $\sqrt{3}$, it turns out that the average log-ratio between female and male employees across all combinations of age groups and types contract is -0.176.  
An important source of information are odds ratio coordinates, which for the Czech Republic result in 
$$(z_{11}^\mathrm{rc}, z_{11}^\mathrm{rs}, z_{12}^\mathrm{rs}, z_{11}^\mathrm{cs}, z_{12}^\mathrm{cs})'=(-0.965, -0.249, 0.391, -0.528, 1.128)'.$$ 
Coordinate $z_{11}^\mathrm{rc}$ compares the type of contract of male and female employees,
and the negative value indicates that the proportion of males with full-time contract, 
compared to those employed on part-time, is higher than the same proportion of females or, alternatively, that the proportion of females is higher within employees with a part-time contract than within those with a full-time contract. The raw odds ratio between these four groups (formed by geometric mean across all age groups) equals 0.33 and the mean log-odds ratio across the age groups is -1.12.
Coordinates $z_{11}^\mathrm{rs}$ and $z_{12}^\mathrm{rs}$ compare the age structure of 
male and female employees and complete the information carried by 
the balances $z_1^\mathrm{r}, z_1^\mathrm{s}, z_2^\mathrm{s}$. 
The coordinate $z_1^\mathrm{s}$ reveals that the youngest group (15--24) is dominated 
by older employees -- the coordinate $z_{11}^\mathrm{rs}$ adds that this dominance tends 
to be slightly higher for male employees. On the other hand, the value of the coordinate $z_{12}^\mathrm{rs}$ indicates that the dominance of the age group 25--54 over 55+ tends to be higher for females. Also the 
coordinates $z_{11}^\mathrm{cs}$ and $z_{12}^\mathrm{cs}$ can be interpreted in the 
sense of odds ratios, by comparing the proportion between full- and part-time contracts in several age groups. The last group of coordinates is formed by 
$z_{111}^\mathrm{rcs}$ and $z_{112}^\mathrm{rcs}$, for the Czech Republic with values $0.124$ and $-0.310$, respectively. These coordinates inform about mutual relations between all three factors and their interpretation becomes a bit tricky. Despite of this complexity 
(comparable to the complexity of double interaction terms in regression models), the 
interpretation in the sense of a double odds ratio is still possible. For instance, 
it was already derived that females are employed more often part-time than males; due to a positive value of $z_{111}^\mathrm{rcs}$ it can be concluded that this relation differs according to the age of employees, specifically it becomes less visible in the youngest 
group. The proposed system of orthonormal coordinates is appropriate for the further statistical analysis of the relations within each cube. For a more detailed interpretation, 
the function {\tt cubeCoordWrapper} also allows to compute all coordinates without the normalising constant and therefore to easier quantify the respective relations.

\subsection{Statistical Analysis}

\begin{figure}[htp]
	\centering
	\includegraphics[width=1\linewidth]{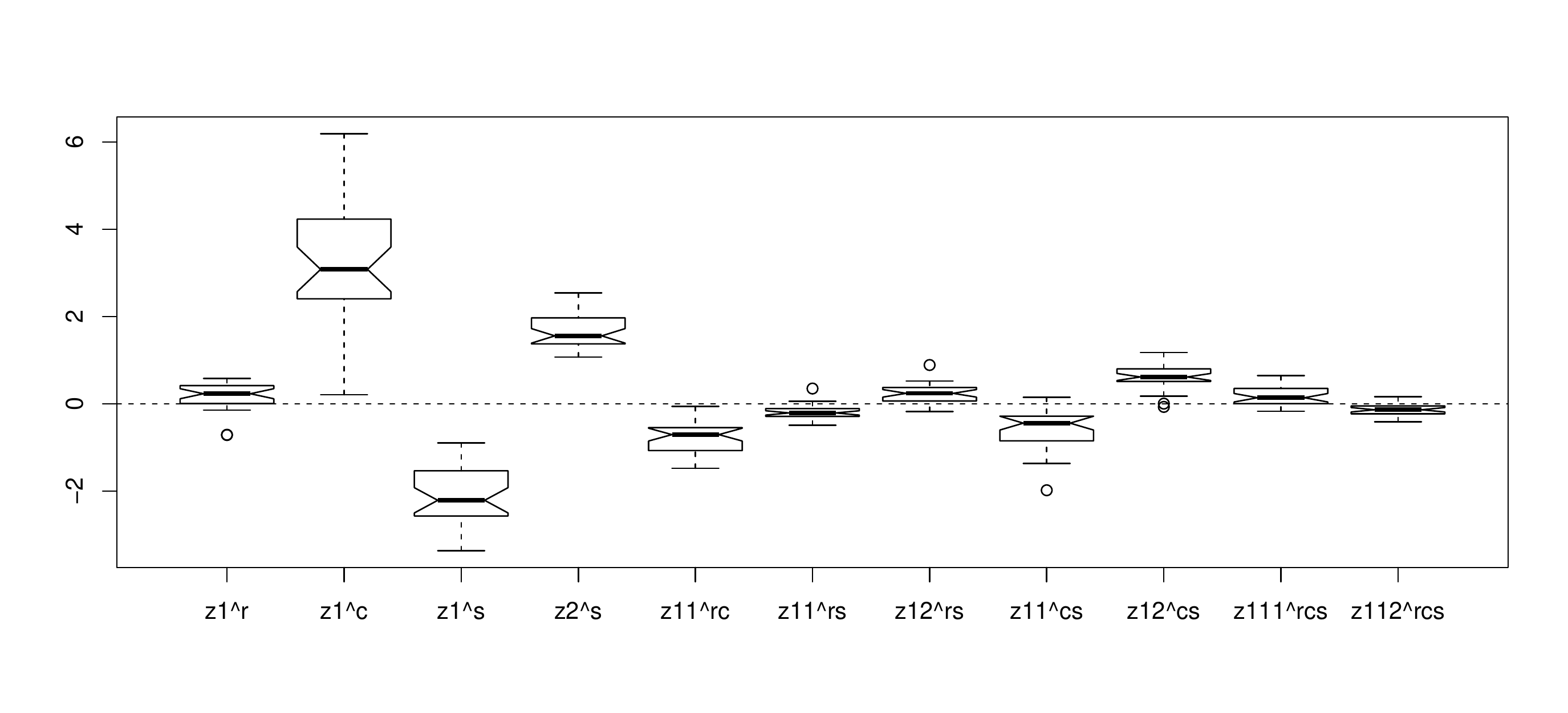}
	\caption{Boxplots of the coordinates describing the employment structure.}
	\label{fig:boxplots}
\end{figure}

Since a sample of 32 compositional cubes is available, this sample can be investigated
in the light of the relative structure.
Due to the geographical and economical proximity of some countries, the assumption of independence of the observations seems not to be sufficiently met in this case, 
and even though the proposed coordinate system is in general designed to allow for any statistical 
processing, this prevents from using standard inference here. First of all, the behavior of the coordinates 
in the sample can be described using boxplots, see Figure~\ref{fig:boxplots},
and $95\%$ bootstrap confidence intervals for the means (both computed by {\tt cubeCoordWrapper}), 
which are collected together with the sample mean values and standard deviations in Table~2.

\begin{table}[htp]
	\label{tab:bootCI}
	\begin{center}
		\begin{minipage}{\textwidth}
			\caption{List of sample means, standard deviations, and $95\%$ bootstrap confidence intervals for the mean of the coordinates describing the employment structure.}
			\begin{tabular*}{\textwidth}{@{\extracolsep{\fill}}lrrclrrc@{\extracolsep{\fill}}}
				\toprule
				&mean&sd&CI&&mean&sd&CI\\
				\midrule
				$z_1^\mathrm{r}$& 0.171 & 0.322 & (0.064, 0.271)&
				$z_{11}^\mathrm{rs}$&-0.182& 0.164 & (-0.235, -0.124)\\
				$z_1^\mathrm{c}$& 3.246 & 1.289 & (2.849, 3.697)&
				$z_{12}^\mathrm{rs}$& 0.230& 0.217 & (0.158, 0.307)\\
				$z_1^\mathrm{s}$&-2.102 & 0.638 & (-2.323, -1.903)&
				$z_{11}^\mathrm{cs}$&-0.591& 0.490 & (-0.752, -0.426)\\
				$z_2^\mathrm{s}$& 1.666 & 0.411 & (1.538, 1.809)&
				$z_{12}^\mathrm{cs}$& 0.631& 0.286 & (0.537, 0.724)\\
				$z_{11}^\mathrm{rc}$&-0.812& 0.333 & (-0.919, -0.701)&$z_{111}^\mathrm{rcs}$& 0.179& 0.222 & (0.108, 0.255)\\
				&&&&	$z_{112}^\mathrm{rcs}$& -0.134& 0.135 & (-0.182, -0.087)\\
				\bottomrule
				
			\end{tabular*}
		\end{minipage}
	\end{center}
	
\end{table}

The scale of the different coordinates presented as boxplots in 
Figure~\ref{fig:boxplots} is comparable since it always refers to the log-ratios of
the employment data. It is obvious that the countries in the sample differ mainly in 
the coordinate $z_1^\mathrm{c}$, comparing the proportionality of full-time and part-time contracts. 
Larger differences are also visible in coordinate $z_1^\mathrm{s}$, where negative
values for all countries are obtained, and thus employees older than 25 years dominate.
According to the bootstrap confidence intervals
shown in Table~2, the effects of the different factors and 
factor combinations represented by the coordinates are all significant.
Thus, not only the previously mentioned simple balances between the factor levels but also interactions between factors strongly influence the overall employment structure.

\begin{figure}[ht!]
	\centering
	\begin{subfigure}{.5\textwidth}
		\centering
		\includegraphics[width=.95\linewidth]{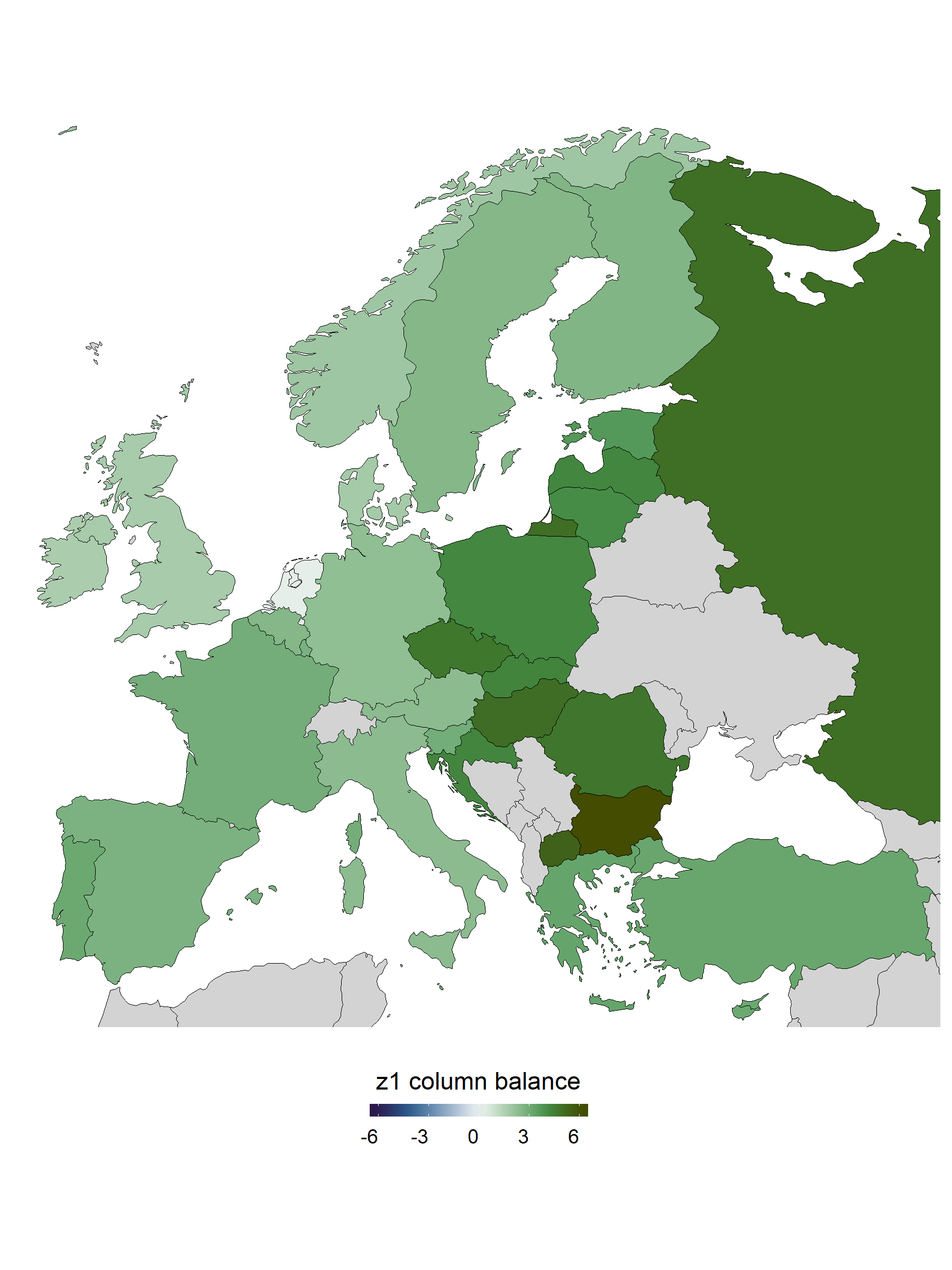}
	\end{subfigure}%
	\begin{subfigure}{.5\textwidth}
		\centering
		\includegraphics[width=.95\linewidth]{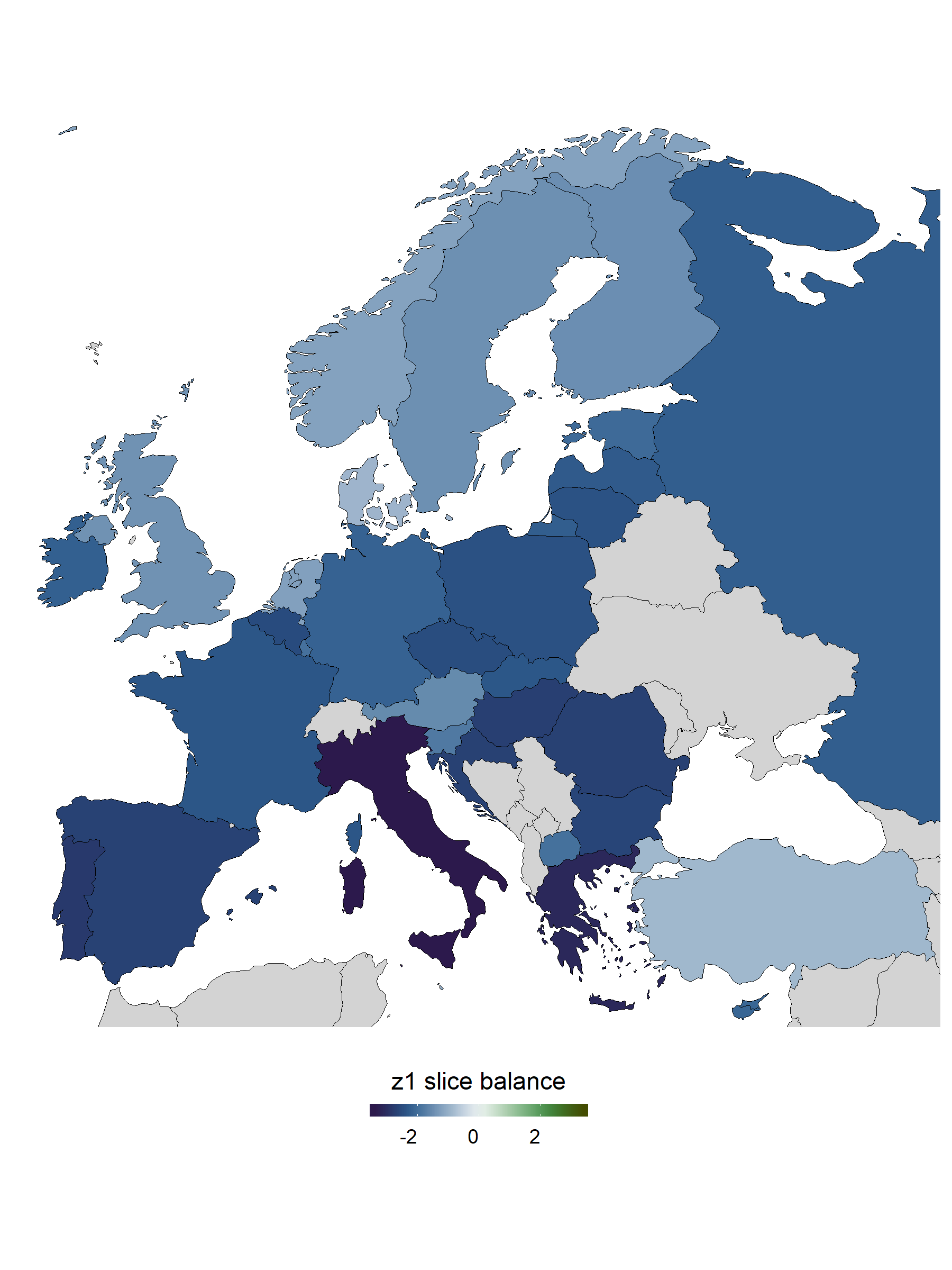}
	\end{subfigure}
	\caption{Values of the coordinate $z_1^\mathrm{c}$ representing the log-ratio between full-time and part-time contracts (left) and the coordinate $z_1^\mathrm{s}$, which represents the log-ratio between the youngest group of employees and the rest (right).}
	\label{fig:maps1}
\end{figure}

\begin{figure}[ht]
	\centering
	\includegraphics[width=0.5\linewidth]{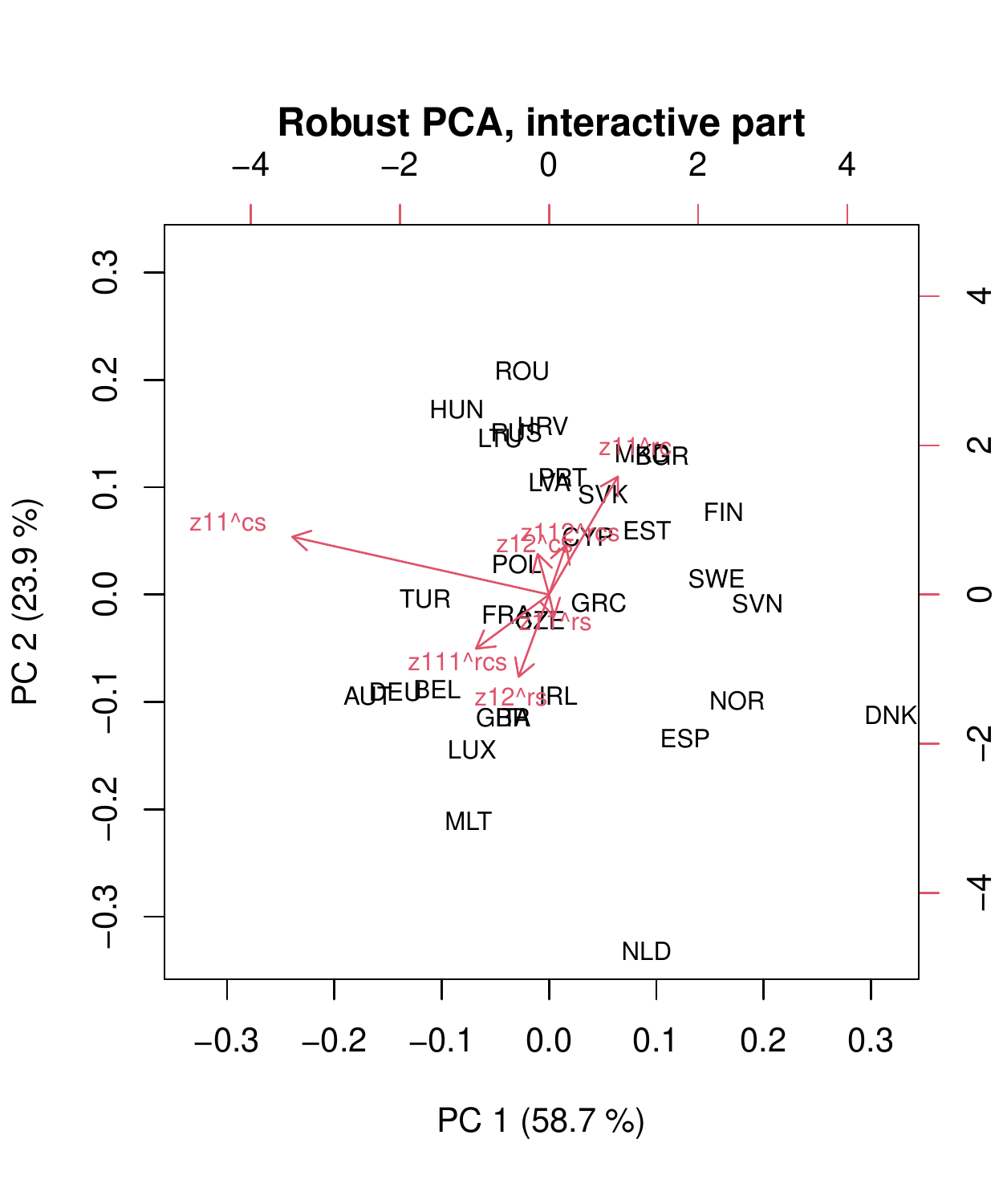}
	\caption{Biplot of the first two principal components resulting from the robust PCA, performed on the coordinates of the interactive part of the compositional cube.}
	\label{fig:biplot}
\end{figure}

Besides these general statements, the coordinate representation also allows for a graphical visualisation of the regional patterns. E.g. values of the most variable coordinates $z_1^\mathrm{c}$ and $z_1^\mathrm{s}$ are shown on Figure \ref{fig:maps1}. From the left map it is clearly visible that high values of coordinate $z_1^\mathrm{c}$, and therefore a big dominance of the full-time contracts, are typical for countries which used to be under the influence of the former Soviet Union. On the other hand, the highest negative values of the coordinate $z_1^\mathrm{s}$, and therefore the biggest prevalence of the older group of employees (25+), are typical for the southern countries like Italy, Greece or Spain.

\begin{figure}[ht]
	\centering
	\begin{subfigure}{.5\textwidth}
		\centering
		\includegraphics[width=.95\linewidth]{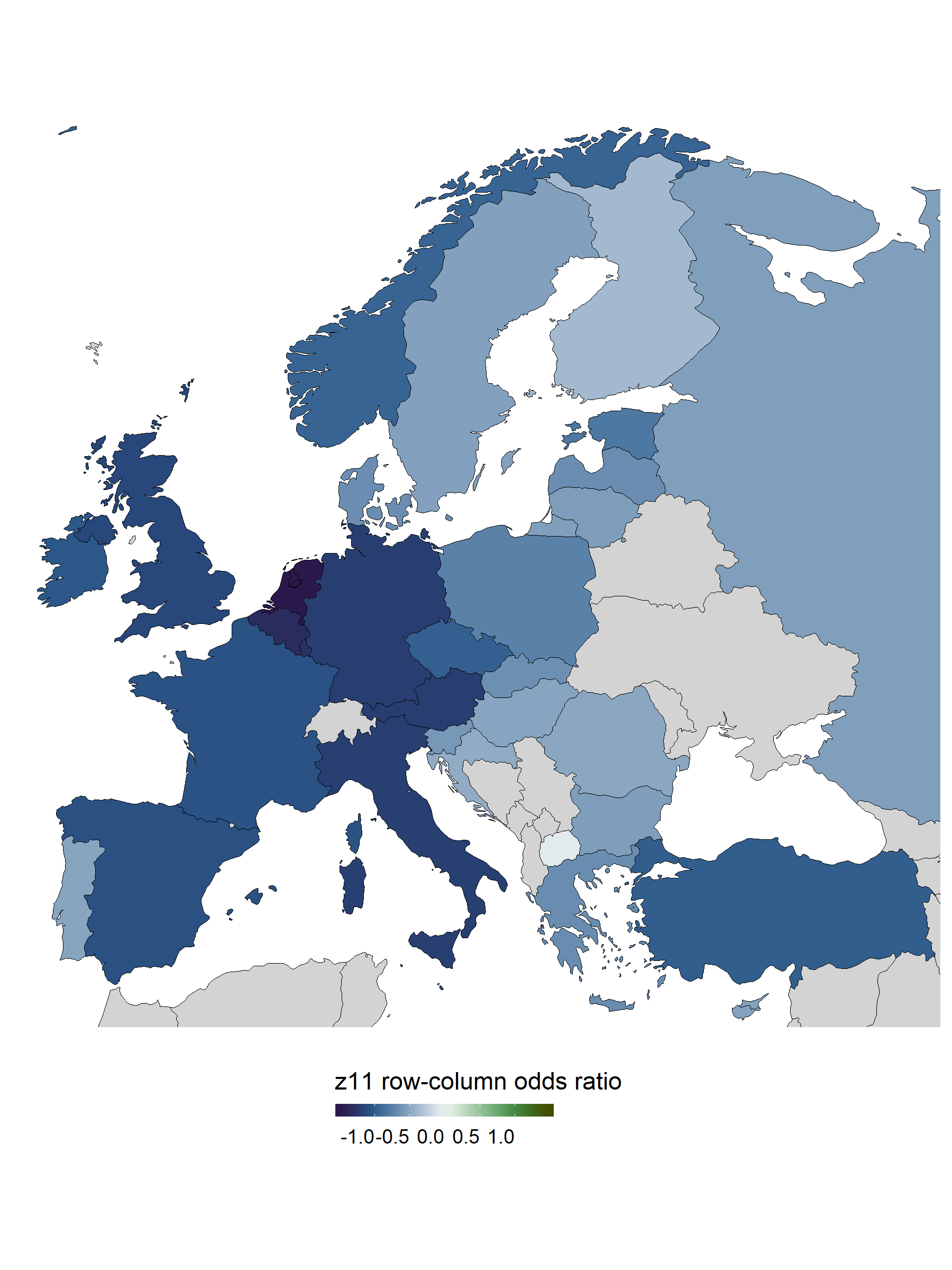}
	\end{subfigure}%
	\begin{subfigure}{.5\textwidth}
		\centering
		\includegraphics[width=.95\linewidth]{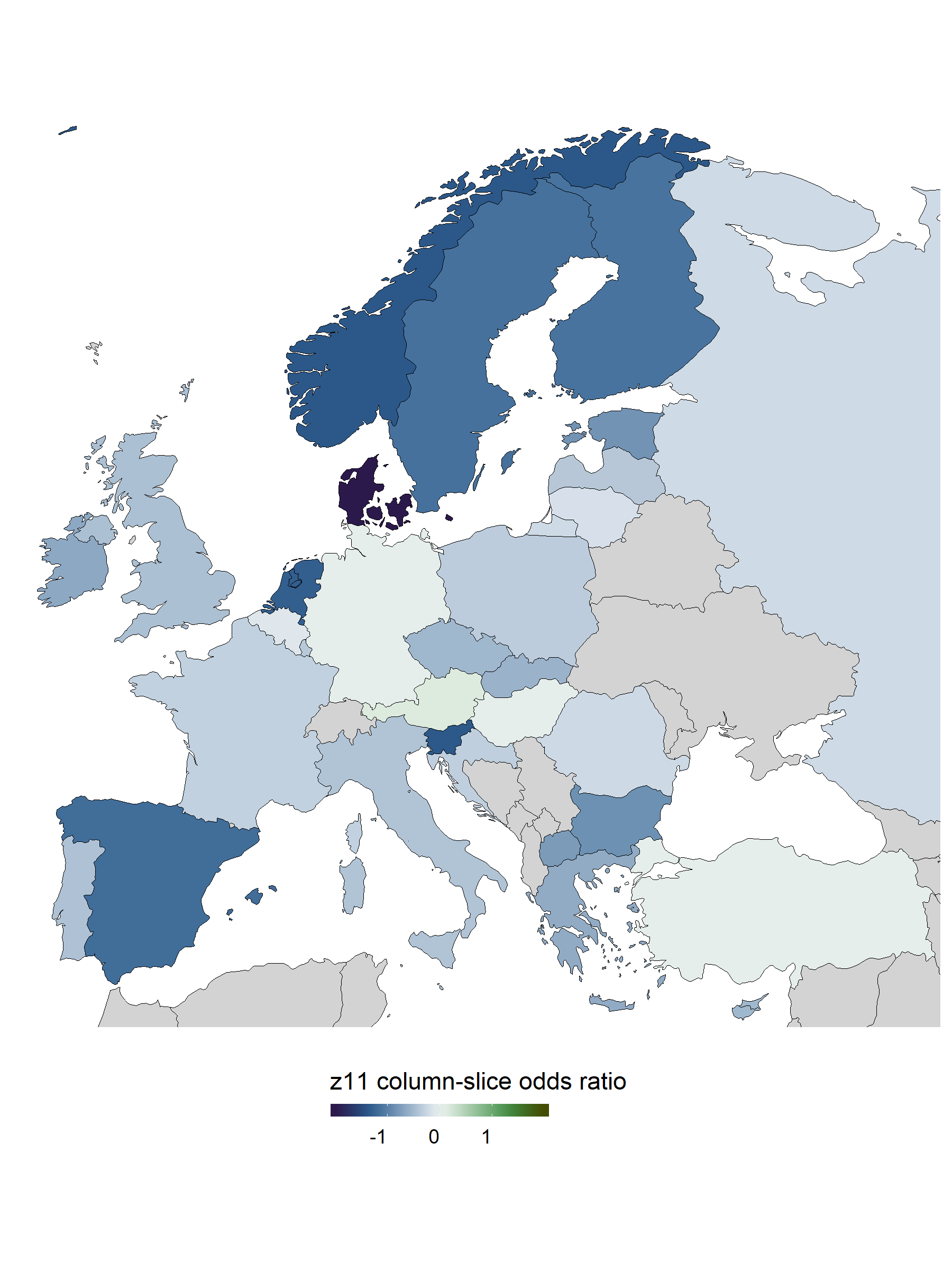}
	\end{subfigure}
	\caption{Values of the coordinate $z_{11}^\mathrm{rc}$, comparing the FT/PT ratio between the female and male employees (left) and the coordinate $z_{11}^\mathrm{cs}$, which compares the same ratio but between the youngest group of employees and the rest (right).}
	\label{fig:maps2}
\end{figure}

Even though the simple balances $z_1^\mathrm{c}$ and $z_1^\mathrm{s}$ already carry an important piece of information about the employment structure, it suppresses the influence of the remaining factors, as described in Section \ref{sec:coord_rep}. The possible deviations from the independence between factors are preserved in coordinates $z_{11}^\mathrm{rc}$ -- $z_{112}^\mathrm{rcs}$ and the main sources of variability in this regard can be found e.g. using principal component analysis (PCA) applied on this set of coordinates. Moreover, since the proposed coordinate system respects the dimensionality of the interactive part of the cube, also a robust version of PCA, based on the minimum covariance determinant (MCD) estimates of location and covariance, can be used. This idea was intensively described for the case of compositional tables in \cite{rendlova18}. The biplot based on the first two robust principal components is shown in Figure \ref{fig:biplot}. According to this result, we can say that the main sources of differences between the countries in the sample, in terms of deviations from independence, are the coordinates $z_{11}^\mathrm{rc}$ and $z_{11}^\mathrm{cs}$. The left map in Figure \ref{fig:maps2} visualises the values of the coordinate $z_{11}^\mathrm{rc}$. 
This coordinate is negative for every country in the sample, and therefore the FT/PT ratio is always higher within male employees compared to females. The biggest differences appear in Netherlands, Belgium, Germany, Austria or Italy; this can be caused e.g. by the popularity of part-time contracts for female employees. 
On the contrary in the countries from the eastern part of Europe, where part-time contracts are not very popular in general (see Figure \ref{fig:maps1}), the difference is less visible. The right map in Figure \ref{fig:maps2} shows values of coordinate $z_{11}^\mathrm{cs}$. Also this coordinate is mostly negative, the FT/PT ratio is therefore higher for the older group of employees (25+). Or, conversely, we can say, that the part-time contracts are mostly popular in the age group 15-24. This difference is mostly visible for Denmark and the Scandinavian countries, followed by Netherlands, Spain and Slovenia. In the remaining countries, the effect of age on the FT/PT ratio is rather negligible.

\begin{figure}[ht]
	\centering
	\begin{subfigure}{.5\textwidth}
		\centering
		\includegraphics[width=.95\linewidth]{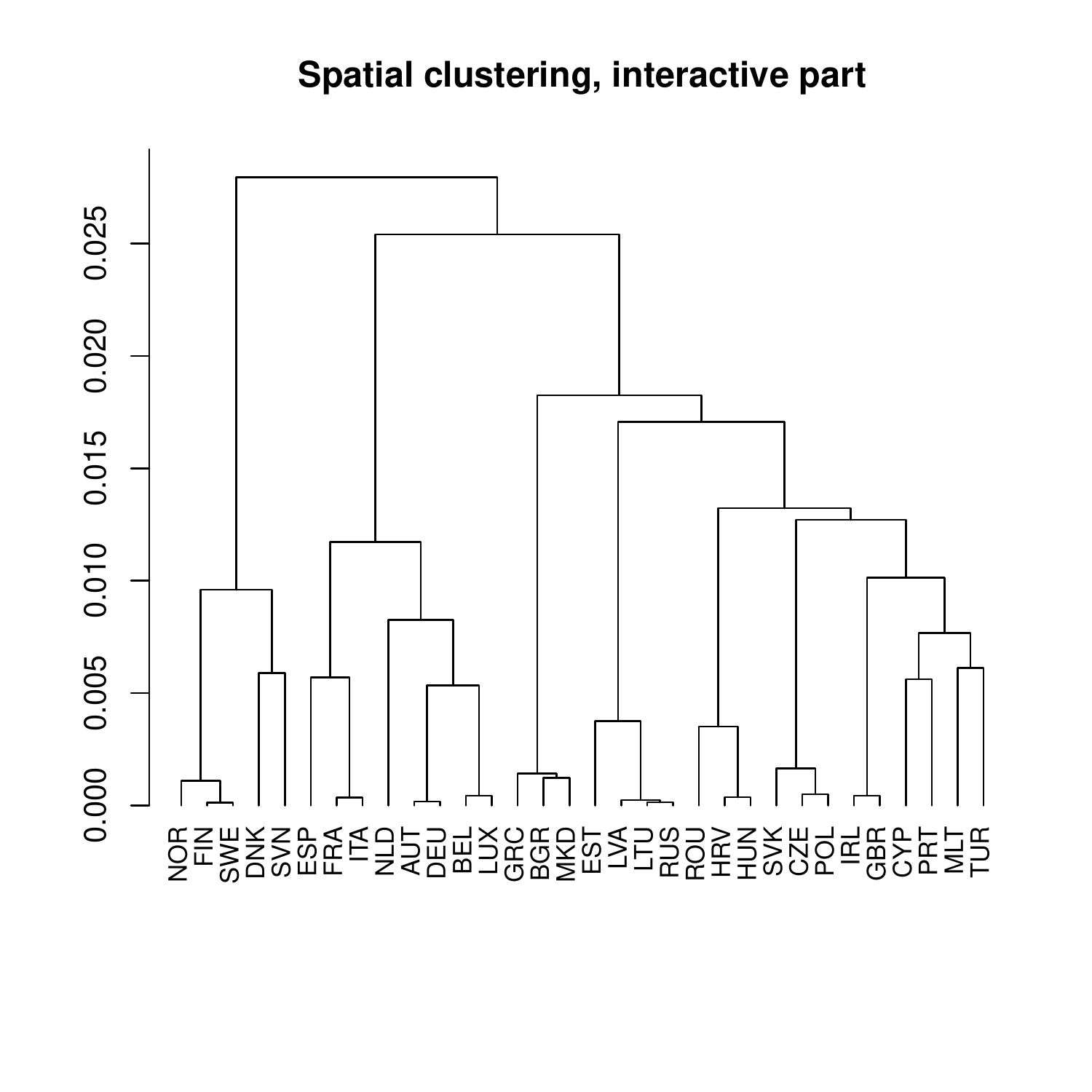}
	\end{subfigure}%
	\begin{subfigure}{.5\textwidth}
		\centering
		\includegraphics[width=.95\linewidth]{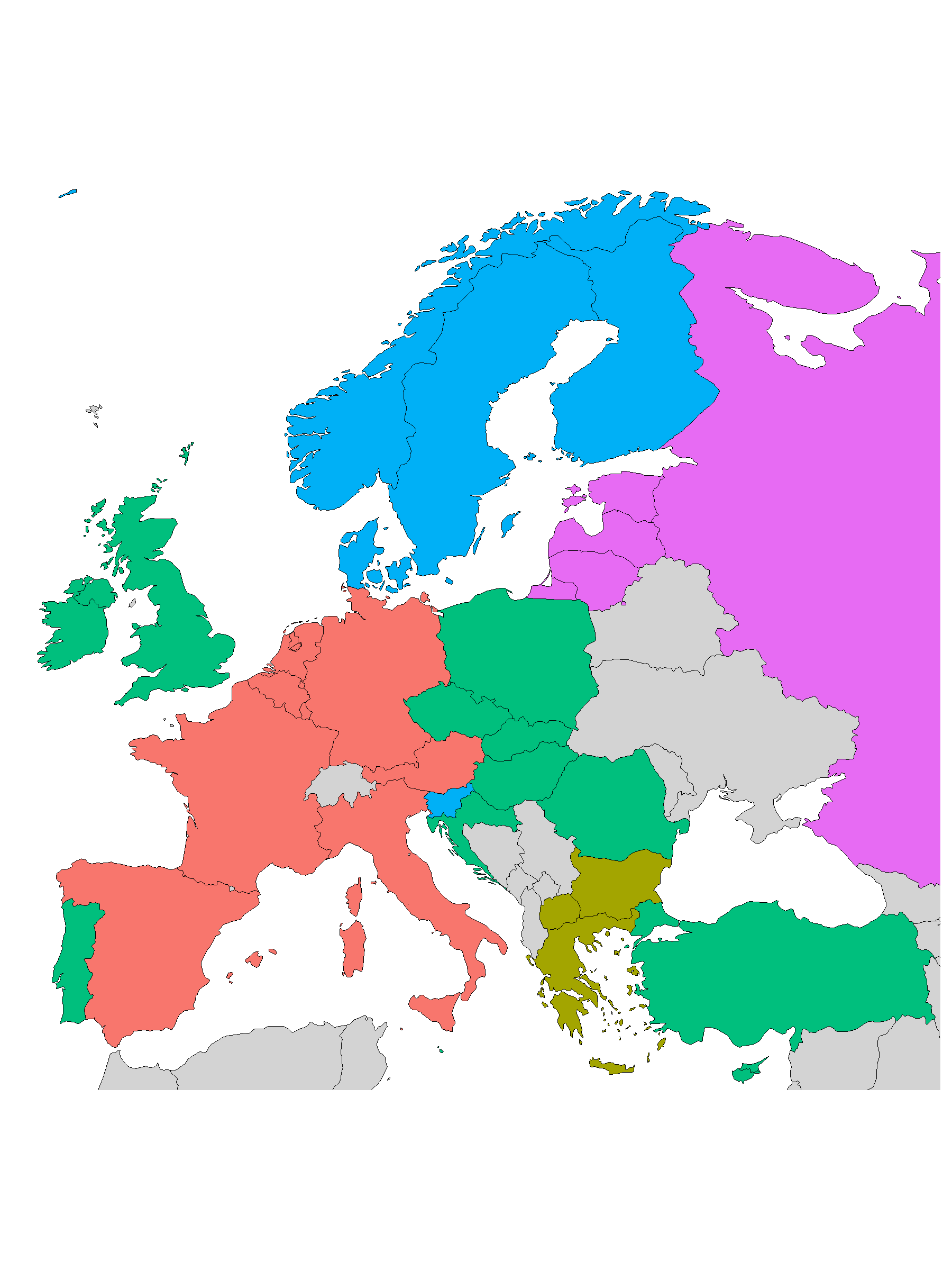}
	\end{subfigure}
	\caption{Result of spatial clustering based on coordinates $z_{11}^\mathrm{rc}$ -- $z_{112}^\mathrm{rcs}$ representing the interactive part of compositional cubes.}
	\label{fig:clustering}
\end{figure}

Robust PCA generates some clusters of countries with similar characteristics in terms of deviations from the independence of the factors. For example, Finland, Sweden, Norway and Denmark have high values on the first component, Romania, Croatia, Bulgaria and Macedonia have high values on the second principal component. A natural question is, whether it is possible to obtain geographically compact clusters of countries with similar employment structure and what are the main features defining these clusters. For this purpose, a clustering method based on two dissimilarity matrices was used \cite{chavent17}. In this method, the first matrix measures the dissimilarity between the numerical variables, in our case the coordinates $z_{11}^\mathrm{rc}$ -- $z_{112}^\mathrm{rcs}$, and the second provides information about neighboring countries. The result is shown in Figure \ref{fig:clustering}. The first cluster is formed by the Scandinavian countries and Slovenia, with high negative values for $z_{11}^\mathrm{cs}$ and very low, (almost zero) values for coordinate $z_{111}^\mathrm{rcs}$. Thus, these countries are characterized by a high popularity of part-time contracts within the youngest group of employees (see the right map on the Figure \ref{fig:maps2} for comparison), which moreover holds despite their gender. The second cluster includes countries from Western and Southern Europe, 
such as Austria, Germany, France, Italy and Spain, with high popularity of part-time contracts for female employees. This property is represented with high negative values
of coordinate $z_{11}^\mathrm{rc}$ and clearly visible in Figure \ref{fig:maps2}. Finally, the Baltic countries and Russia form another compact cluster. They are
characterized by a low difference in the ratio between female and male employees within the middle age (25-54) and the oldest (55+) group ($z_{12}^\mathrm{rs}$ close to zero) and also by a small difference in the same ratio within the employees for full-time and part-time contracts ($z_{11}^\mathrm{rc}$ close to zero).

\section{Example: Mobility Data}
\label{sec:example2}

In the second example, the interest is in the change of the mobility behavior 
of people in Austria within the time period February 3rd until August 2nd, 2020,
thus during the first period of the COVID'19 pandemic. Mobility is measured through the radius of gyration (ROG), a time-weighted distance of the daily movement locations 
of mobile phones to the main location of the phone owner, see \cite{heiler20a, heiler20b} for details. The phone owners are classified with respect to gender and
five age groups (15-29, 30-44, 45-59, 60-74, 75+). The mobility of each group is represented by the respective median value of ROG. This dataset was already analysed in \cite{heiler20a}, where the relative differences in the mobility between the age groups were studied through the clr coefficients. This compositional analysis showed an interesting change in the behaviour of the youngest (15-24) and oldest (75+) part of the population during the lockdown period from March 16th to April 6th, 2020 (weeks 12, 13 and 14). The current results are based on the separate analysis of males and females, when a more complex insight can be reached by a simultaneous study of the age and gender structure. Moreover, when the daily records are aggregated according to the part of the week, the relative differences in mobility over weekdays and weekends can be taken into account. The data at hand can therefore be understood as time series of 26 compositional cubes, each representing the relative mobility structure within one week from weeks
6-31 of 2020. The row levels are formed by gender (F -- female, M -- male), 
the columns by the parts of the week (WD -- weekdays, WE -- weekend), and 
the slices represent the different age groups.

Prior to the main part of the analysis, the SBP of the slice factor needs to be defined. With respect to the findings in \cite{heiler20a}, a separation of the economically active (15-59) and non-active (60+) population seems to be advisable in the first step. The results of the simple clr analysis help to define also the remaining steps of the SBP:
In the second step, the youngest group is separated from those aged between 25 and 59, 
and the third step focuses on the relative dominance of group 30-44 over 45-59. Finally, the relative dominance of mobility within the age group 75+ over 60-74 is highlighted by the last step of the slice SBP. 

\begin{figure}[ht]
	\centering
	\includegraphics[width=1\linewidth]{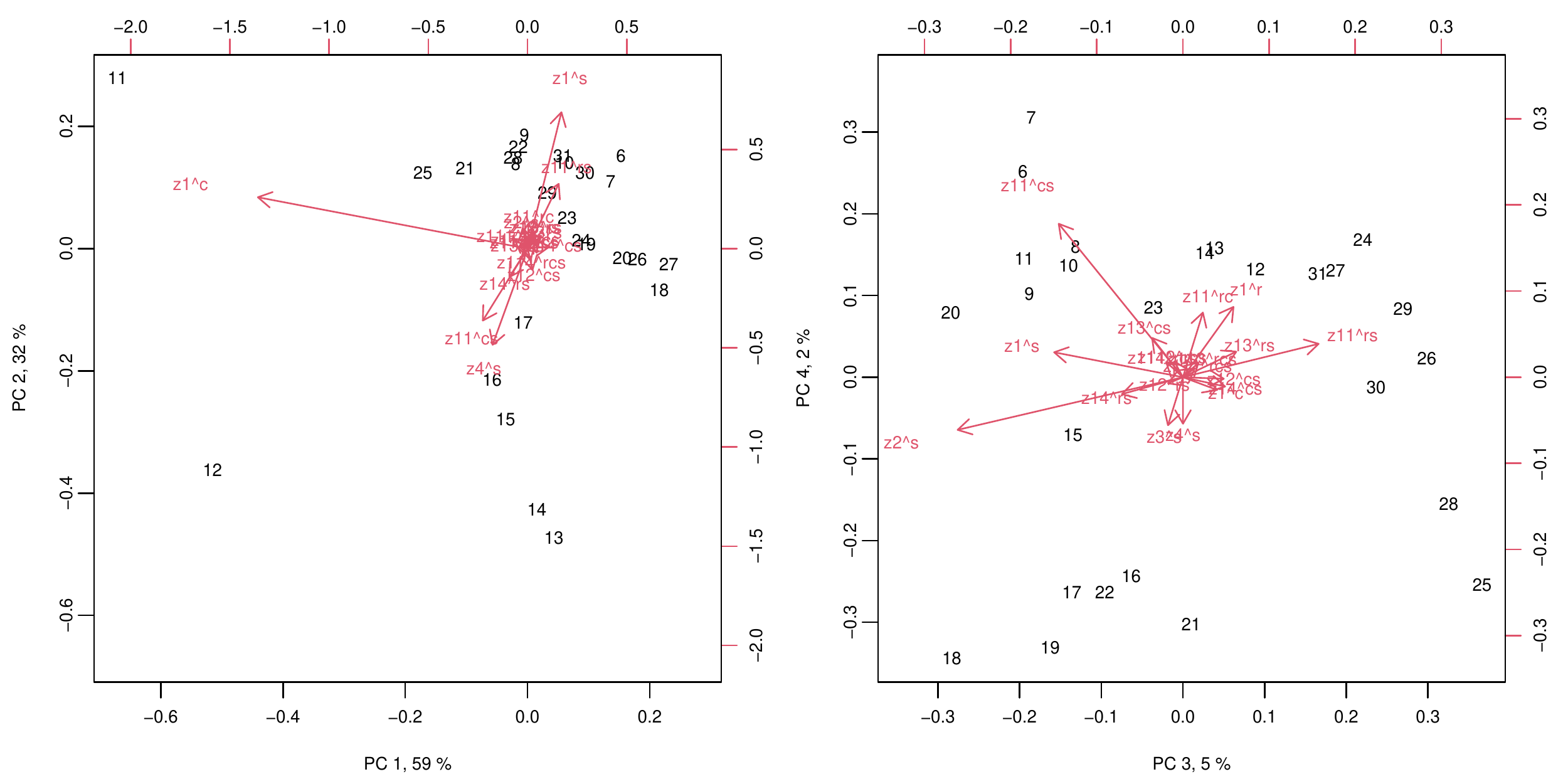}
	\caption{Biplots based on the first four principal components computed for the mobility data. The numbers represent the week numbers in 2020.}
	\label{fig:mobilityBiplot}
\end{figure}

Based on this coordinate representation, some interesting patterns and their sources can be revealed by PCA. Figure \ref{fig:mobilityBiplot} shows biplots based on the first four principal components, describing $98 \%$ of the whole variability.

\begin{figure}[ht]
	\centering
	\includegraphics[width=0.8\linewidth]{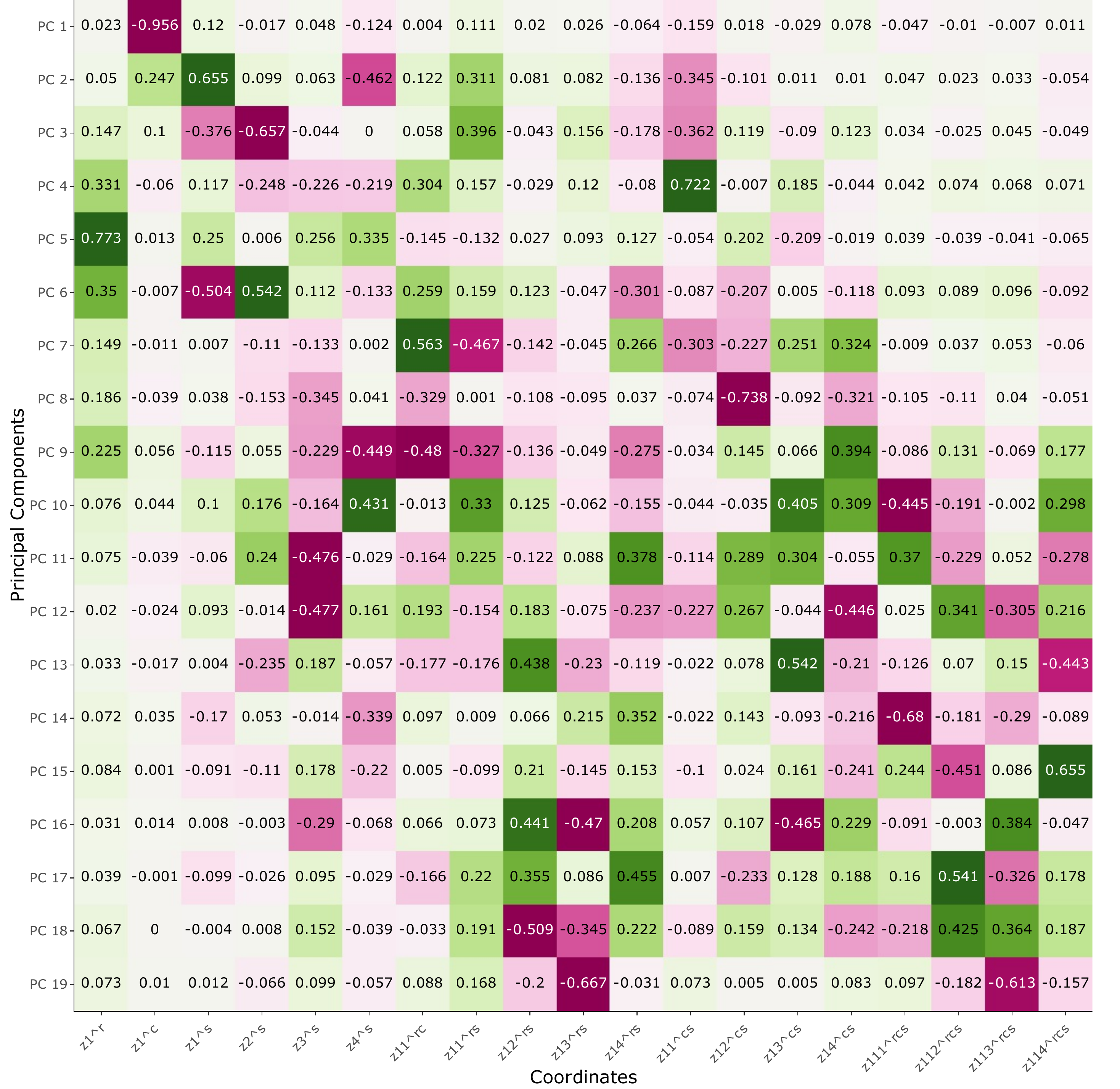}
	\caption{Loadings of the principal components computed for the mobility data.}
	\label{fig:mobilityLoadings}
\end{figure}

The loadings of all PCs are collected in Figure \ref{fig:mobilityLoadings}, 
and it can be seen that the first four principal components are always mainly 
driven by a single coordinate:
\begin{itemize}
	\item PC1 -- $z^c_1$, log-ratio between weekdays and weekend mobility, aggregated over all gender and age groups,
	\item PC2 -- $z^s_1$, log-ratio between mobility of economically active and non-active population, aggregated over all other categories, 
	\item PC3 -- $z^s_2$, log-ratio between mobility of age groups 15-29 and 30-59, aggregated over all other categories, 
	\item PC4 -- $z_{11}^{cs}$, log-odds-ratio comparing economically active and non-active mobility ratio over weekdays and weekends, aggregated over gender.
\end{itemize}

\begin{figure}[ht]
	\centering
	\includegraphics[width=1\linewidth]{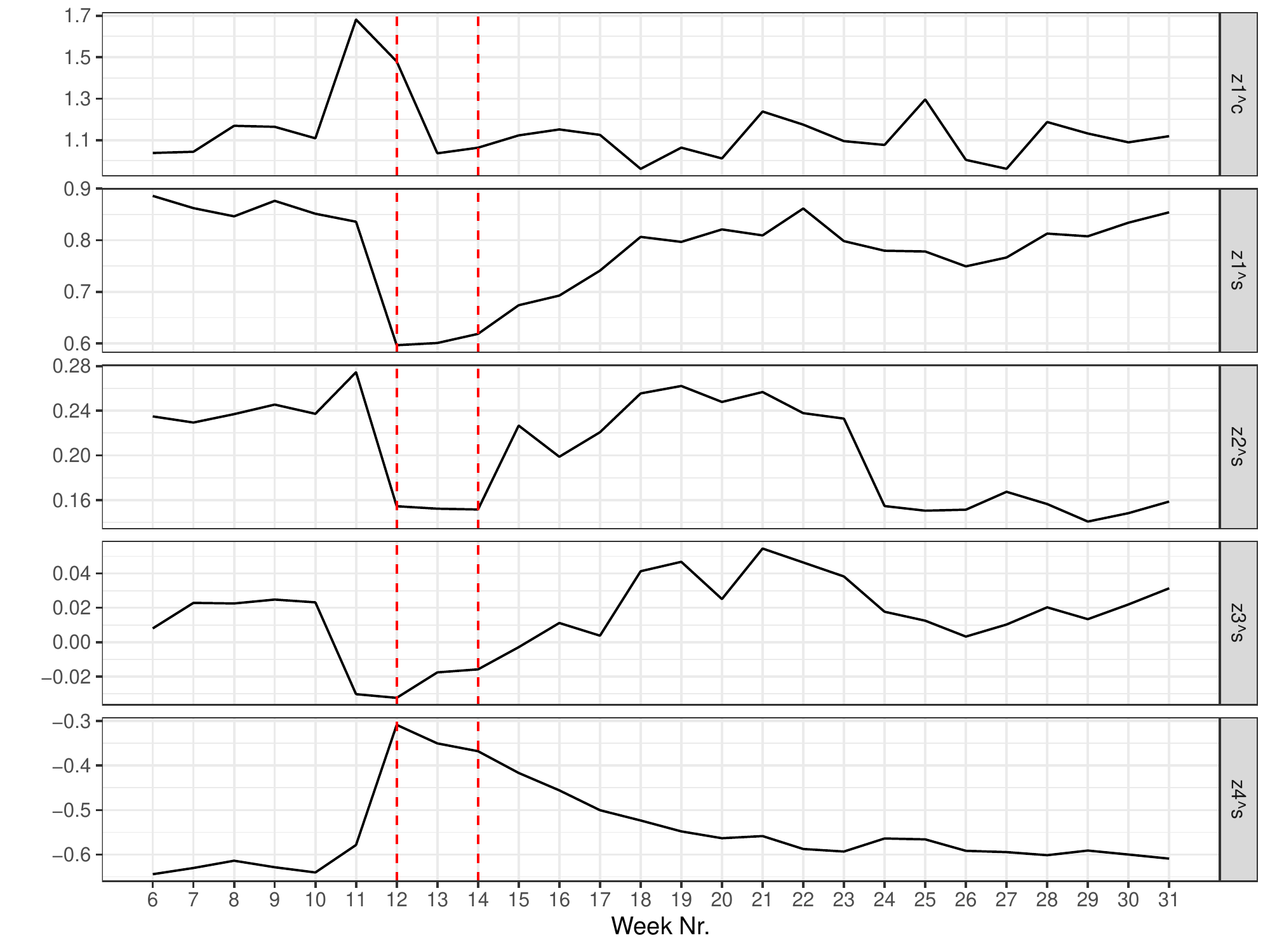}
	\caption{Selected balances (without the normalising constant) computed for the mobility data and their development over time. The red dashed lines define the 
		first lockdown period in Austria.}
	\label{fig:mobilityTS1}
\end{figure}

\begin{figure}[ht]
	\centering
	\includegraphics[width=1\linewidth]{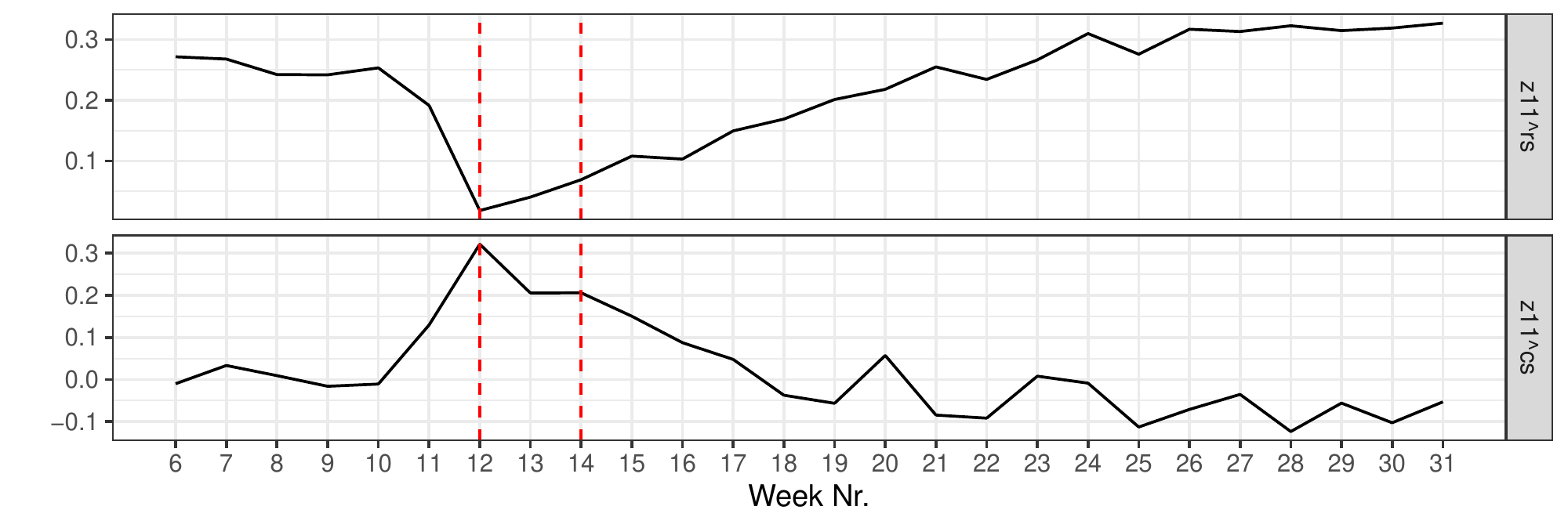}
	\caption{Selected odds-ratio coordinates (without the normalising constant) computed for the mobility data and their development over time. The red dashed lines define the first lockdown period in Austria.}
	\label{fig:mobilityTS2}
\end{figure}

The main source of variability is given by the ratio between weekdays and weekend mobility ($z_1^c$).  As it is visible in Figure \ref{fig:mobilityTS1}, this log-ratio was varying over the whole studied period but it was also atypically high during weeks 
11 and 12 preceding the lockdown period. This gives an evidence on a decrease of the weekend mobility. The second principal component helps to detect typical characteristics 
for the weeks during and immediately after lockdown (weeks 12-17). 
When the relative mobility of the economically active population, quantified by $z_1^s$, was among the lowest during lockdown, mobility of the oldest group 75+ was nearly comparable to the mobility of the population aged between 60 and 74 (coordinate $z_4^s$). Moreover, based on coordinate $z_{11}^{cs}$, Figure \ref{fig:mobilityTS2} 
shows a remarkable difference in the economically active and non-active population mobility ratio during weekdays and weekends, when the former tends to be higher
during lockdown. The weeks from the end of the observed period (weeks 24-31) are 
nicely separated by the third principal component. The relative mobility of the youngest group 15-29 is among the lowest in comparison to the mobility of group 30-59 ($z_2^s$). Moreover, quite stable and high values of coordinate $z_{11}^{rs}$, comparing the ratio 
between the economically active and non-active population for females and males, are typical for this period. Finally, PC4 characterises the behaviour during 
the weeks 16-22 immediately following the lockdown, with the only exception of week 20. According to the respective loadings, the exclusion of weeks 16-22 is not only driven by relatively low values of $z_{11}^{cs}$, but also by a mixture of other phenomenons. For these weeks we observe a high relative dominance of mobility of group 15-29 over 
the mobility of those aged between 30 and 59 ($z_2^s$), and a high relative dominance of the mobility of group 30-44 over group 45-59 ($z_{3}^s$).

\section{Conclusions}

It has been demonstrated that the concepts developed for
two-factorial compositional data can be extended to compositional cubes, and even
to the general $k$-factorial case. The fundamental idea is to 
investigate the
relative data structure in terms of log-ratios between different factors and 
factor levels. One advantage of such an approach is scale invariance, which is
particularly useful when the reported values of the observations are not 
comparable (in our example caused by different population sizes of the countries) 
or if the relative structure is of main interest.

It has been shown that each compositional cube can be decomposed into its independent and interactive 
parts. Furthermore, the interactive part can be decomposed into cubes representing the pairwise factor 
interactions and the interaction between all three factors. It turned out that the components of the interactive part have an advantageous property of uniform marginals and, moreover, that the principle of the decomposition can be directly extended to the general case of $k$-factorial compositions. Since the commonly used systems of orthonormal coordinates are not able to sufficiently describe the multi-factorial nature of cubes and respect the possibility of its decomposition, an alternative system has been proposed. Moreover, this system can be constructed in a flexible manner, basically according
to the needs or expert knowledge of the analyst: There might be a certain hypothesis on the relations
between factors and/or factor levels, and based on the principle of sequential binary 
partitions (SBPs), these combinations can be reflected by the constructed coordinates. Even though the proposed coordinate representation allows to describe the overall relations between 
factors, similar as in the case of the well developed theory for the analysis of vector compositional data,
it is  also possible to use them for
further analysis with standard statistical methods, or to
perform statistical inference with the coordinates, for example,
by constructing bootstrap confidence intervals for the mean, in order to determine
if the effect conveyed by the coordinate is significant. 
A proper coordinate representation of the multi-factorial compositional data can  therefore be understood as a first step in the analysis, possibly followed by
other advanced statistical methods. For example, regression methods with compositional regressors with or without the total \citep{coenders17}, or any other proper methods of one-factorial (vector) compositional data analysis \citep{pawlowsky15} can be used 
(after its possible adaptation to the more complex structure of coordinates).

The idea of modeling interactions between factors using the normalized Hada\-mard product of 
vectors, derived from SBPs at the single factor level, works equivalently for a higher 
number of factors. Therefore, the approach presented here for compositional cubes can be extended in a straightforward manner to higher-order arrays.

\textbf{Acknowledgments}

The work was supported by the Czech Science Foundation, project 22-15684L, and by the Austrian Science Foundation, project I 5799-N.

\textbf{Declaration}

The research of KF and KH leading to these results was supported by the Czech Science Foundation, project 22-15684L and work of PF was supported by the Austrian Science Foundation, project I 5799-N.
All authors contributed to the final version of the manuscript. Kamila Fačevicová developed the main part of the presented theory and prepared the first example. Karel Hron and Peter Filzmoser contributed to the theoretical developments and prepared the second illustrative example. The first draft of the manuscript was written by Kamila Fačevicová and all authors commented on previous versions of the manuscript. All authors read and approved the final manuscript.
The data analysed within the first example are freely available at the OECD data repository http://stats.oecd.org. The mobility data analysed in the second example are not publicly available because of data ownership reasons.

\bibliographystyle{unsrt}  


\end{document}